\documentclass[twocolumn,nobalancelastpage,aps, 10pt, floatfix]{revtex4}  
\usepackage{amssymb}
\usepackage{amsmath}
\usepackage{graphicx}
\usepackage{bbm}
\usepackage{color}
\usepackage{comment}
\usepackage{textcomp, gensymb}
\usepackage{graphicx}
\usepackage{tikz}
\usepackage{hyperref}
\usepackage{float}

\definecolor{blue}{rgb}{0,0,1}
\definecolor{grey}{rgb}{0.6,0.6,0.6}

\newcommand{\bra}[1]{\langle #1 |}
\newcommand{\ket}[1]{| #1 \rangle}
\newcommand{\expval}[1]{\langle #1 \rangle}

\begin{document}

\title{Qubit efficient quantum algorithms for the vehicle routing problem on NISQ processors}

\author{Ioannis D. Leonidas$^{1,3}$}
\email[]{ileonidas@tuc.gr}

\author{Alexander Dukakis$^2$}

\author{Benjamin Tan$^2$}
\email[]{b.tan@u.nus.edu}

\author{Dimitris G. Angelakis$^{1,2,3}$}
\email[]{dimitris.angelakis@gmail.com}

\affiliation{$^1$ School of Electrical and Computer Engineering, Technical University of Crete, Chania, Greece 73100}
\affiliation{$^2$ Centre for Quantum Technologies, National University of Singapore, 3 Science Drive 2, Singapore 117543}

\date{\today}


\begin{abstract}
The vehicle routing problem with time windows (VRPTW) is a common optimization problem faced within the logistics industry. In this work, we explore the use of a previously-introduced qubit encoding scheme to reduce the number of binary variables, to evaluate the effectiveness of NISQ devices when applied to industry relevant optimization problems. We apply a quantum variational approach to a testbed of multiple VRPTW instances ranging from 11 to 3964 routes. These intances were formulated as quadratic unconstrained binary optimization (QUBO) problems based on realistic shipping scenarios. We compare our results with standard binary-to-qubit mappings after executing on simulators as well as various quantum hardware platforms, including IBMQ, AWS (Rigetti), and IonQ. These results are benchmarked against the classical solver, Gurobi. Our approach can find approximate solutions to the VRPTW comparable to those obtained from quantum algorithms using the full encoding, despite the reduction in qubits required. These results suggest that using the encoding scheme to fit larger problem sizes into fewer qubits is a promising step in using NISQ devices to find approximate solutions for industry-based optimization problems, although additional resources are still required to eke out the performance from larger problem sizes.

\end{abstract}


\maketitle

\section{Introduction}

A well studied class of optimization problems is the Quadratic Unconstrained Binary Optimization (QUBO) problems. The importance of QUBO problems stems from the many industry-wide optimization problems that can be re-formulated into QUBO form \cite{vikstaal2020applying, Exxon, kochenberger2014, glover2018tutorial,salehi2022unconstrained,dominguez2023encoding, mattesi2023financial, Glover2022, Streif2021, venturelli2015quantum, adelomou2020using}, one of which is the vehicle routing problem \cite{dantzig1959truck, braekers2016vehicle, toth2002vehicle}.  
Finding optimal ways to send vehicles on the shortest routes not only saves a company fuel cost while maximizing their available resources (e.g. number of available vehicles), but also brings a positive impact to our environment due to the decreased carbon footprint of an operation \cite{lin2014survey, moghdani2021green, kara2007energy}. 

Finding optimal solutions to QUBO problems remain a challenge due to their NP-hard complexity \cite{F_Barahona_1982}. For typical industrial applications, the vehicle routing problem involves numerous vehicles and possible routes, and enumerating through all possible solutions to find the optimal one is computationally unfeasible. 
Recent advancements in quantum computing have allowed for tools to be developed allowing near-term quantum methods to find approximate solutions to QUBO problems \cite{Bharti2021, sawaya2022encoding}, such as Quantum annealing \cite{FINNILA1994343, Kadowaki1998QA} and variational methods such as the Quantum Approximate Optimization Algorithm (QAOA) \cite{Farhi2014QAOA1,Farhi2014QAOA2, Farhi2019, blekos2023review, hadfield2019quantum, Zhou2018} or hardware efficient parameterized quantum circuits \cite{mohammadbagherpoor2021exploring, chatterjee2021variational, Pablo2021, Egger9222275, amaro2022filtering}. However, the limited capability of state-of-the-art hardware within the era of Noisy Intermediate Scale Quantum (NISQ) devices poses a challenge when applying quantum algorithms to problem sizes typical of applied industry problems \cite{Preskill2018NISQ, leymann2020bitter, lau2022nisq, Bharti2021}.
These traditional approaches require as many qubits as there are classical variables, making it unfeasible to solve industry problems of any meaningful size. 

In aid of this, efforts have gone into reducing the number of binary variables required to solve these optimization problems. Existing approaches include attempts to reformulate the classical problem reduce the number of binary variables required \cite{GLOVER2018829, Lewis2017, Boros2006, Lange2019, Ferizovic2020, Hammer1968}, divide-and-conquer approaches that break down the problem or quantum circuit into smaller parts \cite{shaydulin2023qaoa, Pelofske_2023, Lowe_2023, saleem2022divide, bechtold2023investigating, Peng_2020, Fujii2022, Mitarai_2021, Bravyi_2016, Zhang2022, liu2022hybrid, boost2017partitioning, tang2022scaleqc, piveteau2023circuit}, and qubit encoding schemes  \cite{glos2020space, latorre2005image, Plewa_Sieńko_Rycerz_2021, Tan_2021},  amongst others \cite{fuller2021approximate, teramoto2023quantumrelaxation, rancic2023, winderl2022comparative, Liu_2019, decross2022qubitreuse, hua2023exploiting}.

This work explores the use of the encoding scheme in \cite{Tan_2021}, to reduce the number of qubits required to solve classical binary optimization problems. By using a different method of mapping the qubits to the binary variables, classical binary optimization problems involving \(n_c\) classical variables can now be solved with up to $\mathcal O(\log n_c)$ qubits using gate based quantum devices. This reduction in qubits needed means quantum devices now have the potential to solve larger industry-scale QUBO problems, paving the way for industrial applications on NISQ devices. 

We attempt to investigate whether the use of the encoding scheme allows for NISQ devices to find suitable approximate solutions when applied to industry-relevant QUBO problems, in particular, the Vehicle Routing Problem with Time Windows (VRPTW). We begin by showing how the VRPTW can be formulated as a QUBO, following the route-based formulation outlined in \cite{Exxon}, and introduce how the encoding scheme in \cite{Tan_2021} can be used to reduce the number of qubits required for QUBO problems. Using a testbed of problems consisting of 11, and 16-route problem instances, we compare the solutions obtained using the encoding schemes with the standard mapping of binary variables (i.e. routes in the VRP) to qubits, showing results obtained from both numerical simulation, as well as quantum backends provided by IBMQ, IonQ, and Rigetti via AWS. Lastly, we attempt to push the boundaries on the size of VRPTW problems solved using variational quantum approaches, by using the encoding scheme to fit 128 and 3964-route problems onto a quantum device using only up to 13 qubits and comparing the solutions obtained with the classical commercial solver Gurobi. Using traditional quantum approaches for gate-based devices, this will require more qubits than is currently available.


\section{Vehicle Routing Problem with Time Windows}
\label{sec:VRPTW}
Vehicle Routing Problems (VRP) are a well explored and widely applicable family of problems within logistics and operations and are a generalized form of the well-known Travelling Salesman Problem (TSP) \cite{lenstra1975some}. 
The overarching goal of these problems is to manage and dispatch a fleet of vehicles to complete a set of deliveries or service customers while trying to minimize a total cost. 
Many variations of this problem exist, oftentimes seeking to maximize profits or minimize travel costs \cite{gavish1978travelling, lin2014survey}.

The Vehicle Routing Problem with Time Windows (VRPTW) is a VRP variant that enforces time windows within which individual deliveries must be made \cite{solomon1988survey, solomon1987algorithms, braysy2005vehicle}. Reference.~\cite{Exxon} provides $3$ different methods of mapping the VRPTW to a QUBO, of which we will follow the denser, route-based formulation, requiring fewer binary variables for the same number of destinations as the other formulations.

A problem instance is characterized by a network of nodes $\mathcal{N}$.
This set consists of $N$ nodes representing different customers and an additional $0^{\textrm{th}}$ node, \(d\) to serve as the "depot" node, which must be the initial departure point and final destination for all vehicles \(v \in \mathcal{V} \). 
Nodes are connected by directed arcs, \((i, j) \in \mathcal{E}\), each of which has an associated cost \(c_{ij}\). 
These costs are frequently functions of the travel distance or travel time between two nodes, but other measures can also be used.  
The problem may be specified further by assigning each non-depot node a demand level. For the sake of this work, all vehicles are assumed to be homogeneous both in speed and capacity.

Each node \(i\) has an associated time window \([a_{i},b_{i}]\) and can only be serviced after time \(a_i\) and before time \(b_i\) (although we may permit vehicles to arrive earlier and wait until \(a_i\)). 
The depot can be treated as having a time window of \([0,+\infty]\) and the effective arrival time at node $i_{p+1}$ is given by $T_{i_{p+1}}=\max\{\alpha_{i_{p+1}},T_{i_p}+t_{i_p,i_{p+1}}\}$.

According to this route-based formulation, valid solutions to the problem instance will inform whether or not a route should be travelled.
We define a route \(r\) as an ordered sequence of \(P\) nodes \((i_1,i_2,...,i_P)\) for a vehicle to travel to.
To satisfy the requirement that vehicles must start and end at the depot, we require that \(i_1=i_P=d\).
In order for a route to be considered valid, it must be composed entirely of valid segments: \((i_p, i_{p+1}) \in \mathcal{E}\) \(\forall\) \(1\leq p\leq P-1\).  

We must also ensure that the arrival time \(T_{i_p}\) for any given node \(p\) along route \(i\), must not exceed the upper limit of that node's time window, i.e. \(T_{i_p} \leq b_{i_p} \forall p\).
This set \(\mathcal{R}\) containing all valid routes defines the size of the problem. 
For a route \(r \in \mathcal{R}\) with sequence \((i_1,i_2,...,i_P)\), we can calculate a route cost \(c_r=\sum_{p=1}^{P-1}c_{i_{p},i_{p+1}}\).
We also define a value \(\delta_{i, r}\) which is equal to \(1\) if node \(i\) lies in route \(r\) and \(0\) otherwise. 
This allows us to express the objective of our VRPTW in terms of a vector \(\vec{x}\), containing decision variables \(x_r\) corresponding to each route \(r\):

\begin{equation}
    \underset{\vec{x}}{\textrm{min}} \sum_r^R c_rx_r ,\qquad x_r \in \{0, 1\} \ \forall \ r \in \mathcal{R} \label{Eq:VRPCostFn}
\end{equation}

\begin{equation}
    \textrm{s.t.} \sum_r^R \delta_{ir}x_r = 1 ,\qquad \forall \ i \in \mathcal{N} \label{Eq:VRPConstraints} 
\end{equation}

where \(x_r\) is a binary variable that has value \(1\) if route \(r\) is to be travelled, and 0 if not. 
The linear equality constraint \eqref{Eq:VRPConstraints} ensures that all nodes in the network are visited exactly once within our optimal solution. As an additional optional constraint, it may be desirable to set the number of used vehicles to some fixed value \(V\). This is achieved with the additional condition:
\begin{equation}
    \sum_r^R \delta_{0r}x_r = V \notag
\end{equation}
where we have used the nomenclature of \(\delta_{0,r}\) referring to the inclusion of the depot node in route \(r\). This is included purely for illustrative purposes, as all valid routes must include the depot node and therefore \(\delta_{0,r} = 1 \ \forall \ r \in \mathcal{R}\).

\begin{figure*}
    \centering
    \includegraphics[width=0.9\textwidth]{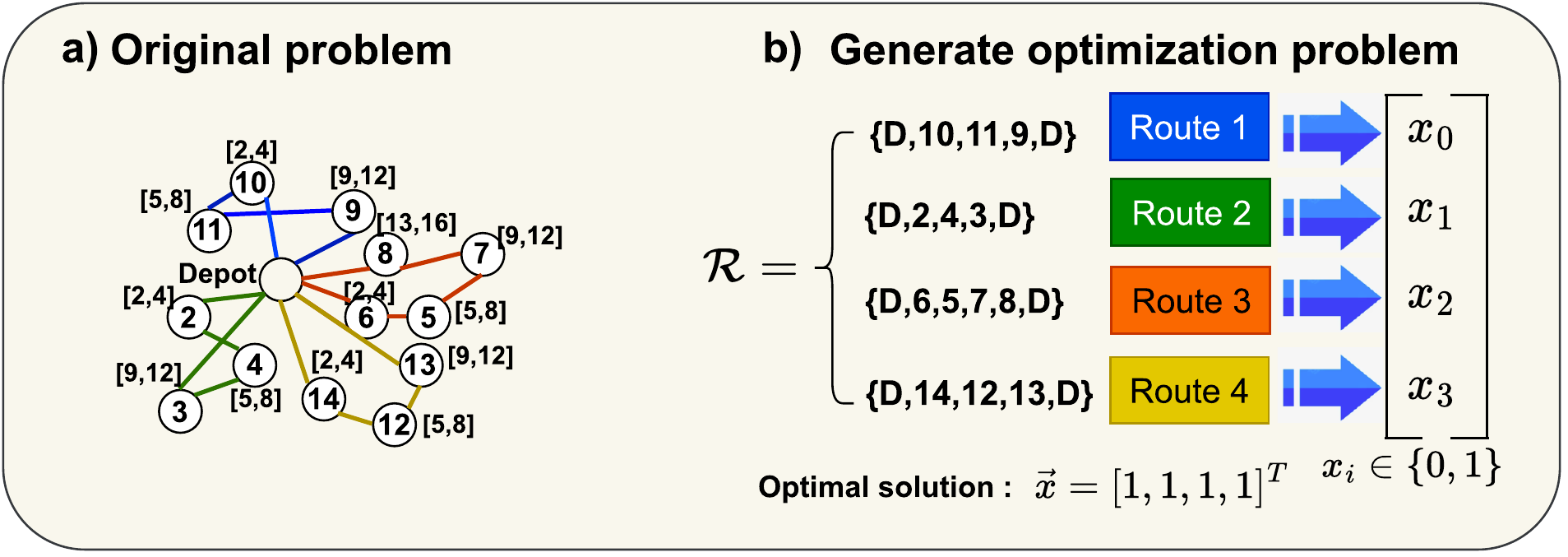}
    \caption{Mapping of a simple 4 route VRPTW optimization problem on a quantum computer using complete and minimal encoding. a) Graph of the original optimization problem. Each node in the graph is associated with a location id and a time window. b) After obtaining the set of feasible routes \cite{Exxon}, we generate the optimization problem by assigning a binary variable to every route. With the minimal encoding we need 3 qubits to represent this 4-route VRPTW problem and the optimal quantum state becomes a superposition of the register qubits followed by the appropriate value of the ancilla qubit e.g. $\ket{\psi}=\frac{1}{2}(\ket{1}\ket{00}+\ket{1}\ket{01}+\ket{1}\ket{10}+\ket{1}\ket{11})$.
    }
    \label{fig:fig_new}
\end{figure*}

A toy example of the VRPTW problem with $14$ destinations without constraints on the number of vehicles is shown in Fig.~\ref{fig:fig_new}a. In this example, we show the destinations being grouped into $4$ routes, with each route starting and ending at the depot. This can be done using a classical heuristic or otherwise, to remove routes that do not respect the time windows or unallowed routes due to other constraints (e.g. untraversable paths, gated communities, long travel time between nodes, etc). Figure.~\ref{fig:fig_new}b shows how these routes are mapped to the binary variables $x_r$, with the optimal solution in this case being all $1$'s as it leads to all the destinations being visited exactly once. 

\section{Quadratic Unconstrained Binary Optimization}

The QUBO problem is a binary optimization problem consisting of finding a binary vector \(\vec{x}^*\) such that:

\begin{equation}
\vec{x}^* = \underset{\vec{x}}{\textrm{arg min}}\, \vec{x}^{\top} \mathcal{A} \vec{x}
\label{Eq:QUBO}
\end{equation} 

where \(\vec{x}\in\{0,1\}^{n_c}\) is a vector of \(n_c\) classical binary variables and \(\mathcal{A}^{n_c\times n_c}\) is a real and symmetric matrix constructed from our optimization problem. 
Classical methods of finding solutions to QUBO problems involve a range of metaheuristic algorithms such as simulated annealing \cite{kirkpatrick1983optimization, Rutenbar17235, Bertsimas1993, Guilmeau9513782, Romeo1991, EGLESE1990271}, TABU search \cite{glover1990tabu, Glover1998, wang2012multilevel, Fouskakis2002} or genetic algorithms \cite{goh2020hybrid, math11010129, Doerr2021} amongst others \cite{Dorigo6787854, Grama755612, Ricardo1996, BLUM2005353, GHAREHCHOPOGH20191, Wang2018, RizkAllah2023}, with many of these having been adapted for the vehicle routing problem \cite{KOULAMAS199441, ELSHERBENY2010123, 7047830, Doerner2010, MONTOYATORRES2015115, Konstantakopoulos2022, Abualigah2020, Cordeau2005, basu2012tabu}.

To find solutions to the QUBO problem using traditional quantum approaches, the cost function is first mapped to an Ising Hamiltonian:

\begin{equation} \label{Eq:HIsing}
\hat H_{\rm \text{Ising}} = \frac{1}{4}\sum_{k,l}^{n_c} \mathcal{A}_{kl}(1-{\hat\sigma}_{z}^{(k)})(1-{\hat\sigma}_{z}^{(l)})
\end{equation}

where \({\hat\sigma}_{z}^{(k)}\) is the Pauli \({Z}\) matrix acting on qubit \(k\), and \(A_{kl}\) are the elements of matrix \(\mathcal{A}\).
The ground state of \({\hat H_{\rm \text{Ising}}}\) is a basis state \(\ket{x_i}\) that corresponds to an exact solution \(\vec{x_i}\) of the QUBO problem defined by \(\mathcal{A}\). 

Traditional implementations of VQAs to search for this ground state maps each binary variable in \(\vec{x}\) to a single qubit, using the same number of qubits as classical variables i.e.~\(n_q = n_c\).
The resulting quantum state is parameterized by a set of angles \(\vec{\theta}\) and can be expressed as a linear superposition of each basis states \(\ket{x_i}\) representing classical solutions \(\vec{x_i}\):

\begin{equation}
	\ket{\psi_c(\vec{\theta})} = {\hat U_c}(\vec{\theta})\ket{\psi_0} = \sum_{i=1}^{2^{n_c}} \alpha_i(\vec{\theta}) \ket{x_i}, \label{Eq:FullEncoding}
\end{equation} 

where \({\hat U_c}(\vec{\theta})\) is the unitary evolution implemented on the quantum computer.
Classically, the cost function for our QUBO problem when minimizing the energy of \({\hat H_{\rm \text{Ising}}}\) can be expressed as:

\begin{equation}
	\textrm{min} \: C_{x} = \sum^{2^{n_c}}_{i=1} \vec{x_i}^{\top} \mathcal{A} \vec{x_i} P(\vec{x_i}). \label{Eq:QUBOprob}
\end{equation} 

$P(\vec{x_i})$ is the probability of sampling the classical solution $\vec{x_i}$. From the quantum state given in \eqref{Eq:FullEncoding}, this probability of sampling the $i^{\rm th}$ solution is given by the $|\alpha_i(\vec{\theta})|^2$.

In section \ref{sec:ExperimentalProcedure}, we describe how the VRPTW formulation as described in section \ref{sec:VRPTW} can be mapped into a QUBO problem of the form given in \eqref{Eq:QUBO}.

\section{Quantum Algorithms for QUBO}

In this section, we outline some of the commonly used methods to solving the vehicle routing problem using near term quantum algorithms. For a more comprehensive review of near term quantum algorithms and their applications to the VRP, we refer our readers to Ref.~\cite{Bharti2021} and \cite{Osaba9781399}.

\subsection{Quantum Annealing}

Quantum annealing is a commonly employed method of solving combinatorial optimization problems using quantum hardware and use cases for vehicle routing problems have been explored \cite{FINNILA1994343,Kadowaki1998QA, AQCReview2018, Yarkoni_2022, Irie2019, harikrishnakumar2020quantum, Borowski2020, Yarkoni2021}. This method starts by having the qubits in the ground state of a Hamiltonian that can be easily prepared, typically \({\hat H_0} = \sum_{i=1}^{n_c} {\hat\sigma}_{X}^{(i)}\), where \(\hat\sigma_{X}^{(i)}\) denotes the Pauli \(X\) gate acting on qubit \(i\). The system is then evolved slowly towards the Hamiltonian that represents the problem. By the adiabatic theorem, the quantum system will arrive in the ground state of the problem Hamiltonian if the evolution is sufficiently slow \cite{AQCReview2018}, and a measurement on the qubits will return the state that minimizes the cost function of the optimization problem at hand. The total Hamiltonian of the system can be expressed as

\begin{equation}
	{\hat{H}} = A(t) \: {\hat{H_0}} + B(t) \: {\hat H_{\textrm{Ising}}}  \label{Eq:QAnnealing}
\end{equation} 

where $A(t)$ and $B(t)$ are functions based off the annealing schedule. $A(0) = 1$ and $B(0) = 0$ at the start of the annealing procedure and $A(t_e) = 0$ and $B(t_e) = 1$ at the end of the annealing schedule $t_e$.

\subsection{Variational hybrid quantum-classical algorithms}

\begin{figure}
    \includegraphics[width=\columnwidth]{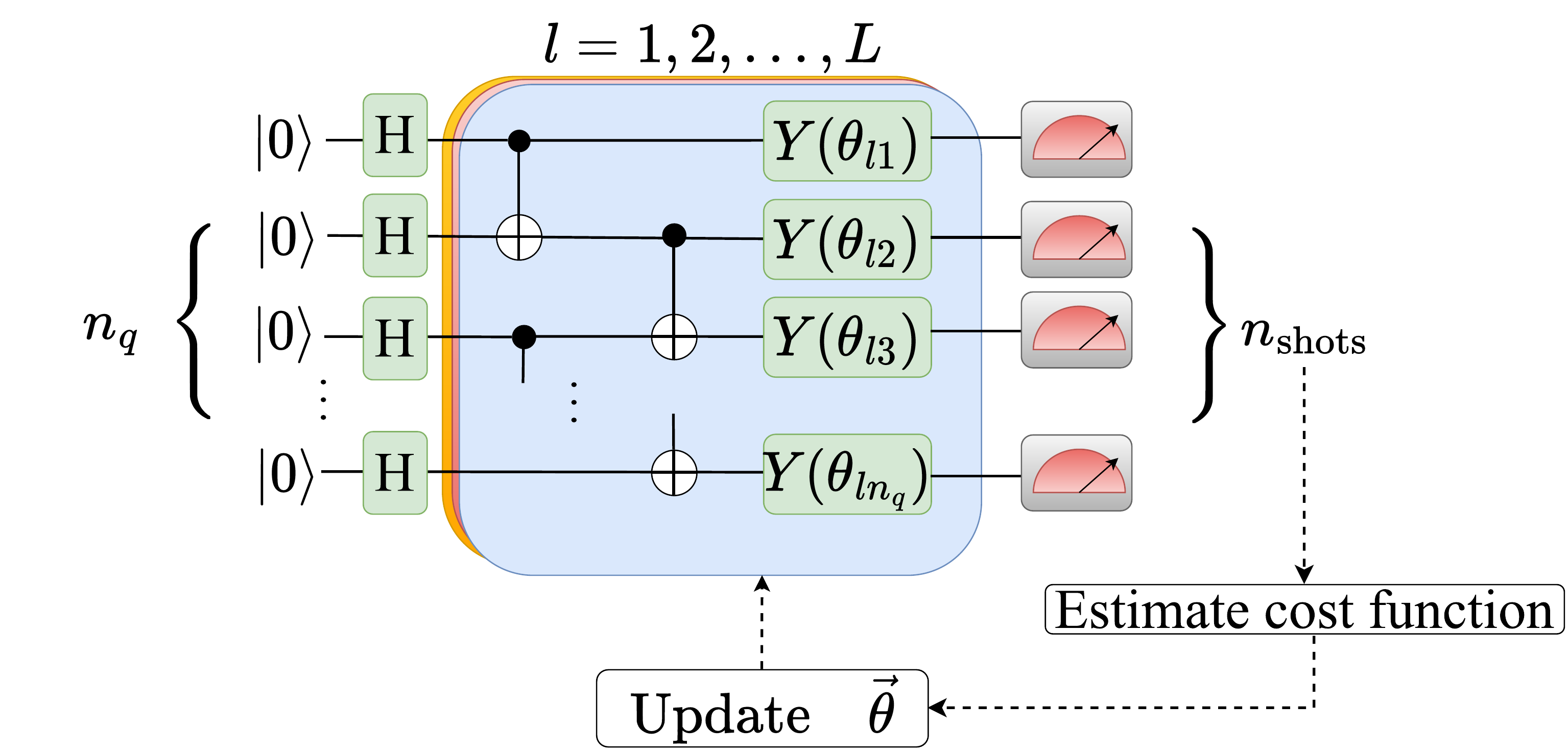}
    \caption{Hardware efficient variational ansatz for all encoding schemes described. Each circuit begins with a layer of Hadamard gates followed by $L$ repetitions of CNOT gates and single qubit $Y(\theta_i)$ rotations. 
The circuit is measured in the computational basis to estimate the respective cost functions, and a classical optimizer is used to update the variational parameters $\vec{\theta}$.}
    \label{fig:circuitansatz}
\end{figure}

Other quantum approaches to solving the VRP include variational methods such as QAOA \cite{Azad9774961, fitzek2021applying, xie2023feasibilitypreserved, Mohanty10214310} and hardware efficient parameterized quantum circuits \cite{Alsaiyari10087522, xie2023feasibilitypreserved}. 
These approaches involve applying some unitary operator $U(\vec{\theta})$ onto some initial quantum state $\ket{\psi_0}$ and using a classical optimizer to adjust the variational parameters $\vec{\theta}$ to find the optimal quantum state $\ket{\psi_{\textrm{opt}}} = U(\vec{\theta}_{\textrm{opt}}) \ket{\psi_0}$.

In the QAOA, this unitary takes on the following form where $P$ is the number of alternating layers of ${U_H} (\gamma) = e^{-i \gamma \hat{H}_{\textrm{Ising}} }$ and $U_X (\beta) = e^{-i \beta \sum_{j=1}^{n_c} \hat{\sigma}_X}$ that are applied to an initial quantum state $\ket{\psi_0} = \ket{+}$ to give

\begin{equation}
	\ket{\psi (\vec{\gamma}, \vec{\beta})} = \prod_p^P U_X(\beta_p) U_H(\gamma_p) \ket{+}. \label{Eq:QAOA}
\end{equation} 

The optimal quantum state is found by adjusting the variational parameters $\vec{\gamma} = [\gamma_1, \ldots, \gamma_P]$ and $\vec{\beta} = [\beta_1, \ldots, \beta_P]$ using classical optimization techniques. 
QAOA, being a trotterization of the annealing procedure into multiple unitaries, is able to provably converge to the optimal solution as $P \rightarrow \infty$ \cite{Farhi2014QAOA1}. However, implementing the QAOA on gate based quantum hardware typically requires deep circuits \cite{kossmann2022deep}. 

In hardware efficient variational approaches, $U(\vec{\theta})$ is constructed using gates that can be efficiently implemented on the specific quantum device available, such as single qubit rotations and two-qubit entangling gates as seen in Fig. \ref{fig:circuitansatz}. 
While easily implemented on quantum hardware, hardware efficient variational approaches typically do not guarantee any convergence to the optimal solution, and suffer from barren plateaus within the cost function landscape, making it difficult to obtain the optimal solution \cite{McClean2018, cerezo2021cost, wang2021noise}.

\begin{figure*}
\centering

\begin{tabular}{cc}
\includegraphics[width=0.5\linewidth]{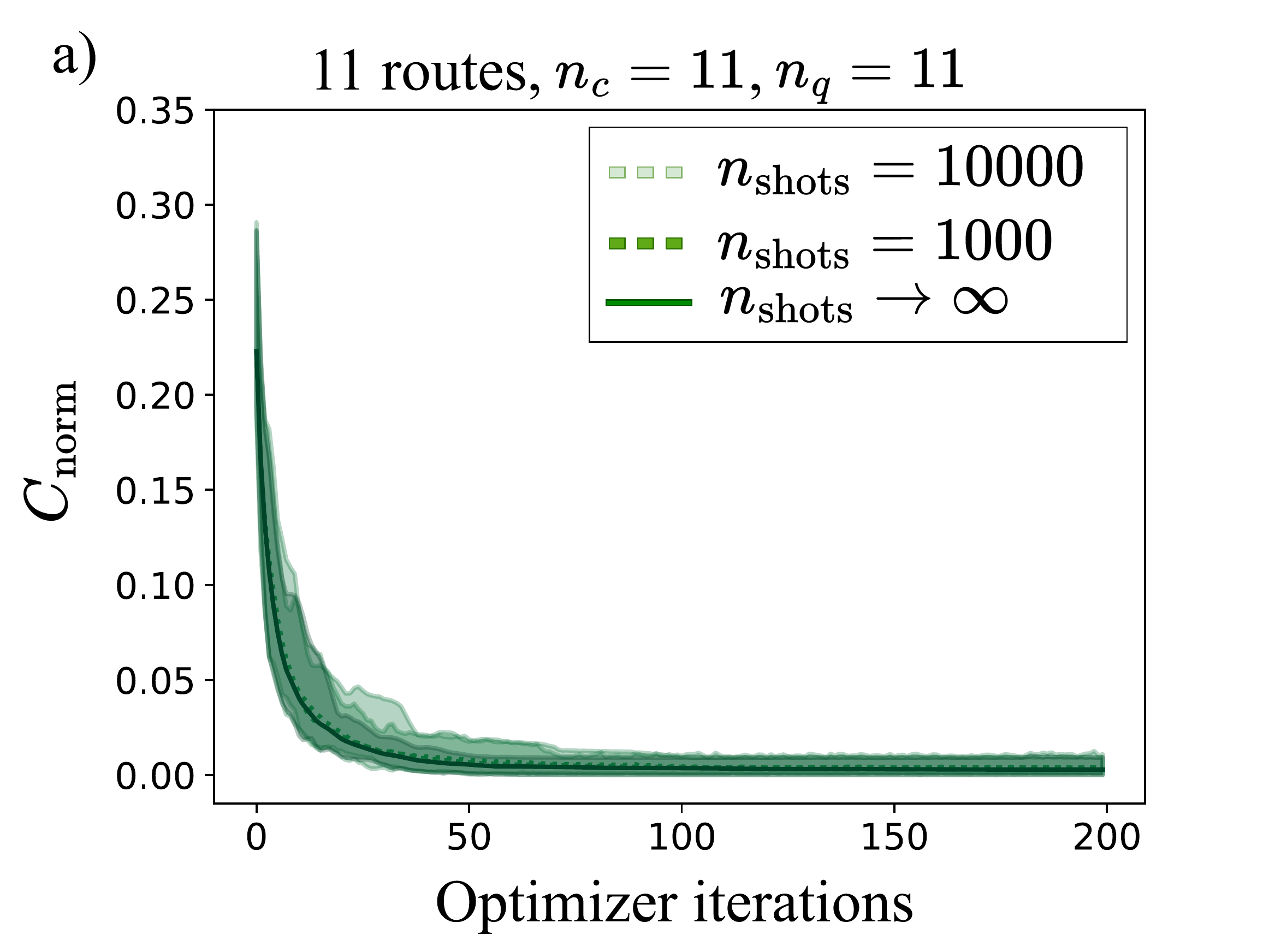}
\includegraphics[width=0.5\linewidth]{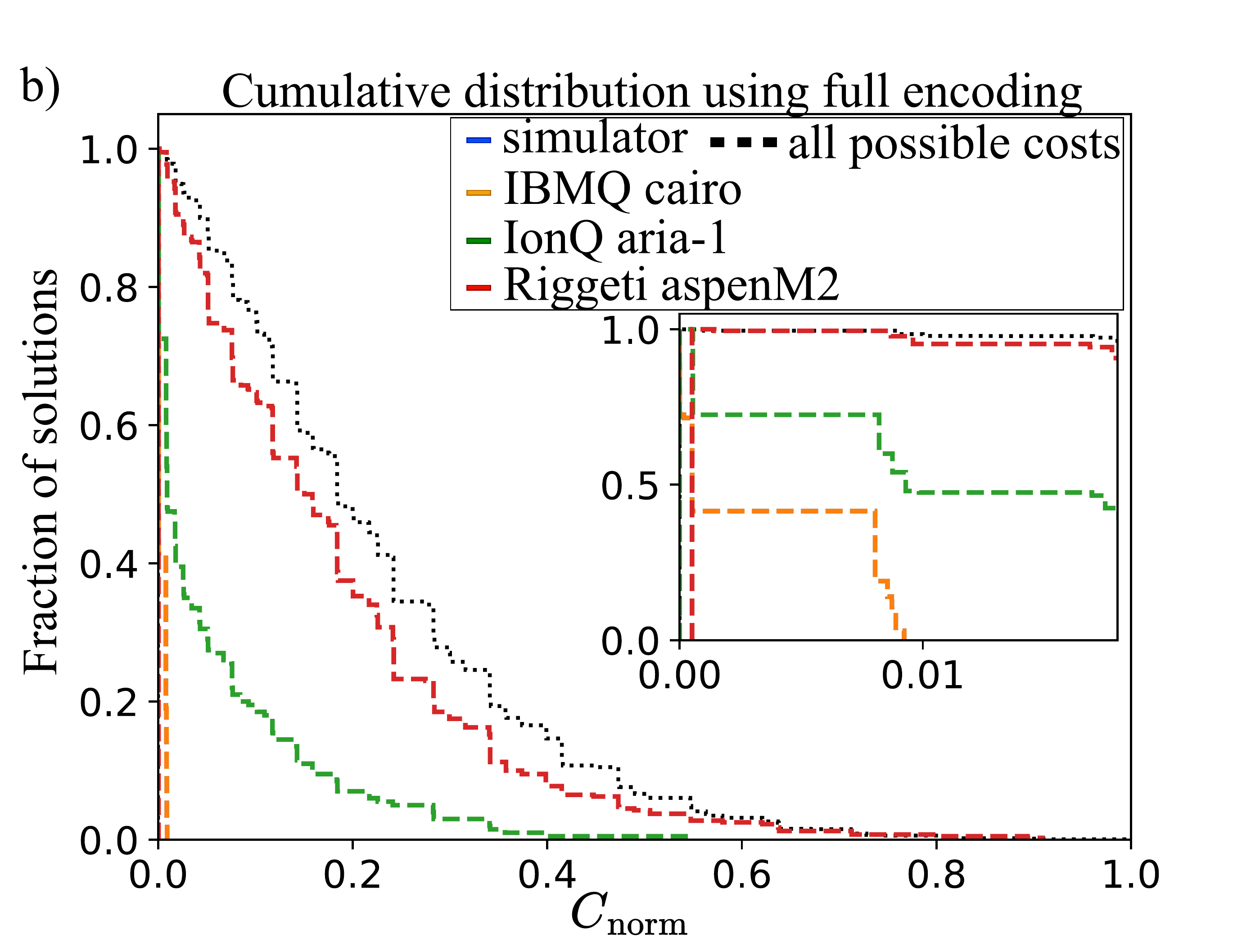}\\
\includegraphics[width=0.5\linewidth]{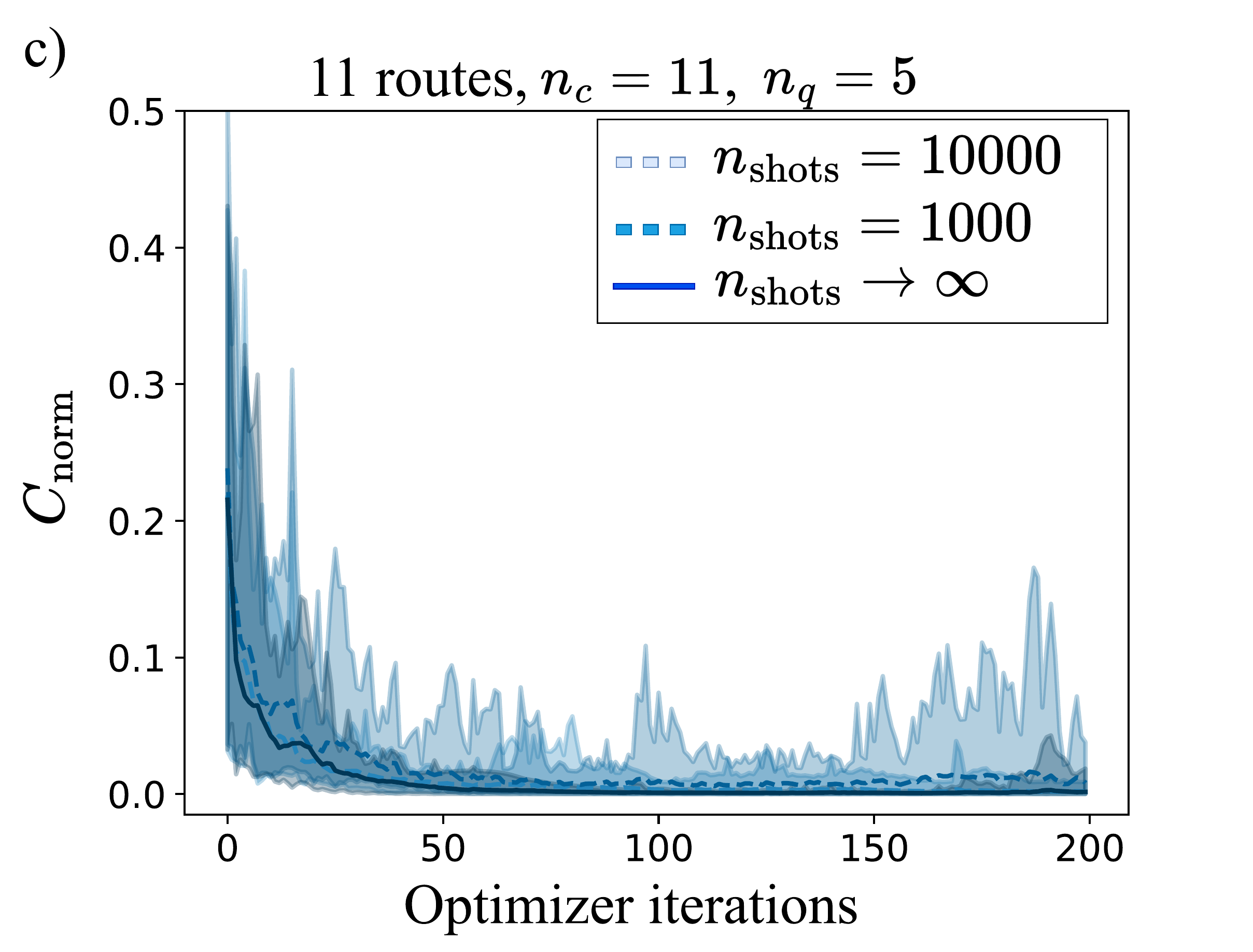}
\includegraphics[width=0.5\linewidth]{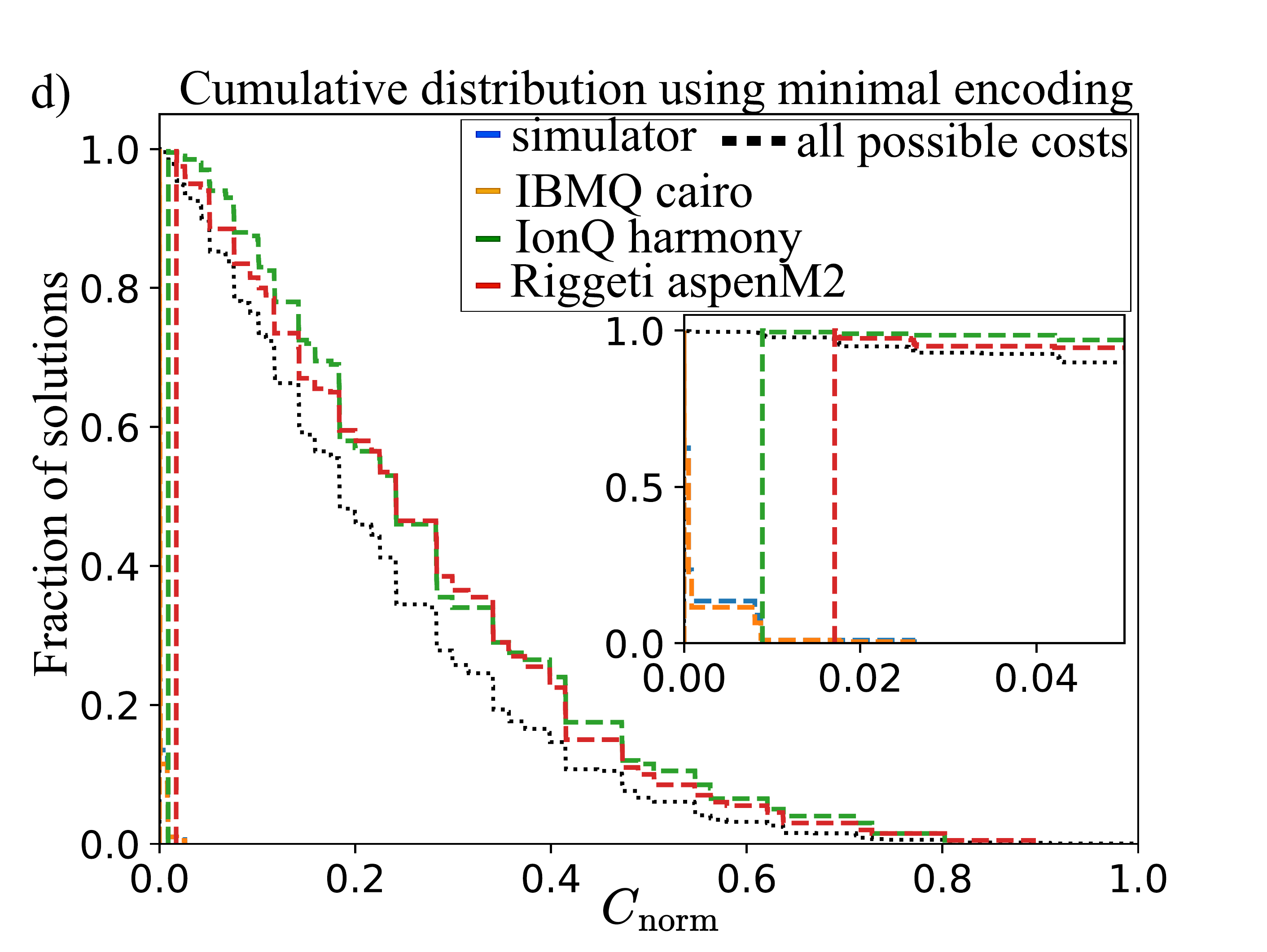}
\end{tabular}
    \caption{Comparison between a ${n_c=11}$ route optimization instance, using full and minimal encoding. 
    a, c) Convergence graphs for the full encoding and minimal encoding schemes with $n_{\textrm{shots}} = 1000, 10000$, and $n_{\textrm{shots}} \rightarrow \infty$.
    Shaded regions show the deviation of the normalized cost function values per optimizer iteration.
    b, d) Cumulative distribution of classical solutions obtained post optimization using full and minimal encoding respectively. Black curves show the distribution of $C_{\textrm{norm}}$ values of all $2^{n_c}$ classical solutions. Blue curves show classical bitstrings obtained using $n_{\textrm{shots}} = 10000$ shots from simulator. Orange, green, and red curves show the classical bitstrings obtained using the optimal parameters from optimization, on IBMQ, IonQ, and on Riggeti quantum backends. Insets show the zoomed version close to  $C_{\textrm{norm}} = 0$.
    }
\label{fig:Results11}

\end{figure*}

\begin{figure*}
\centering

\begin{tabular}{cc}
\includegraphics[width=0.5\linewidth]{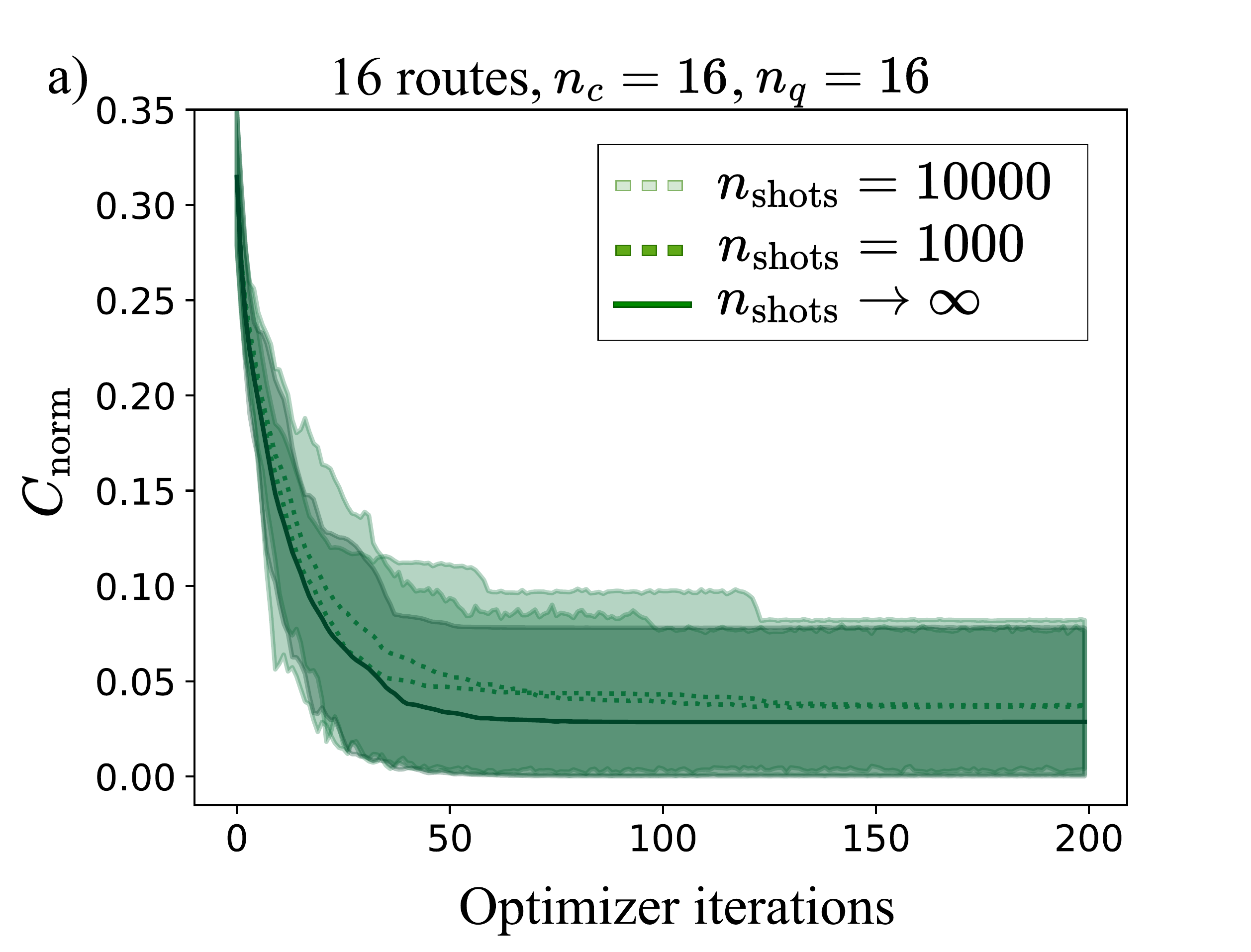} 
\includegraphics[width=0.5\linewidth]{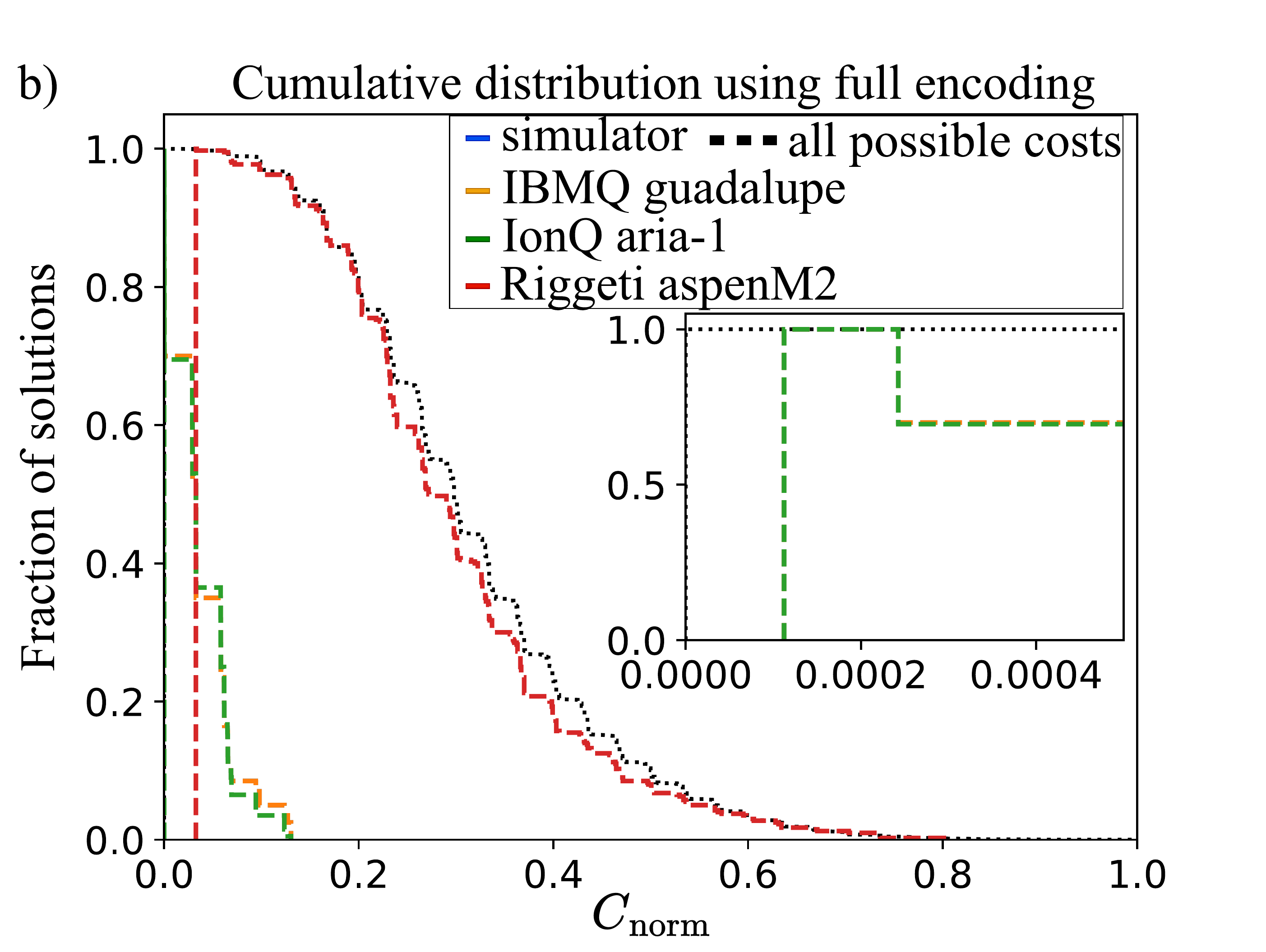}\\
\includegraphics[width=0.5\linewidth]{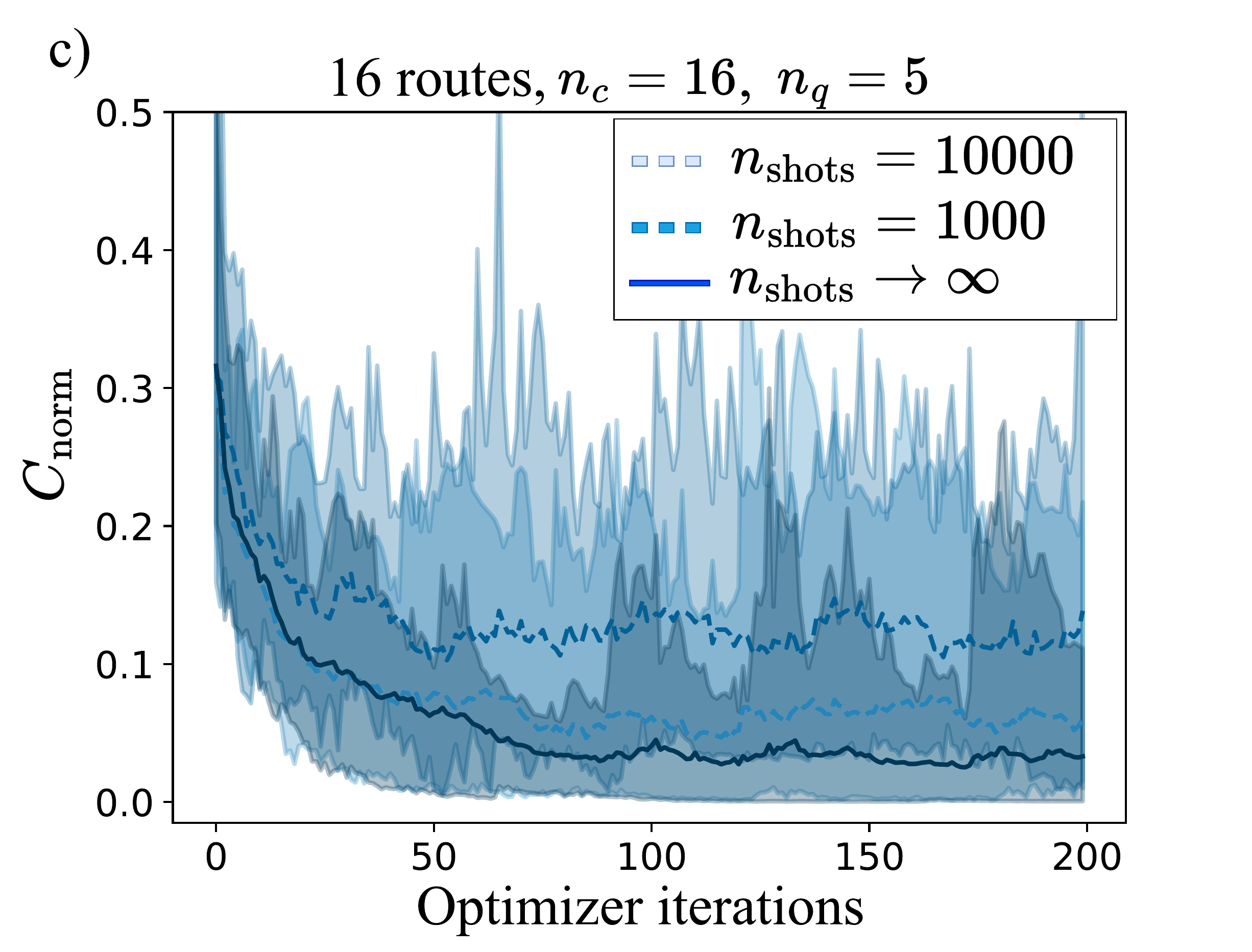}
\includegraphics[width=0.5\linewidth]{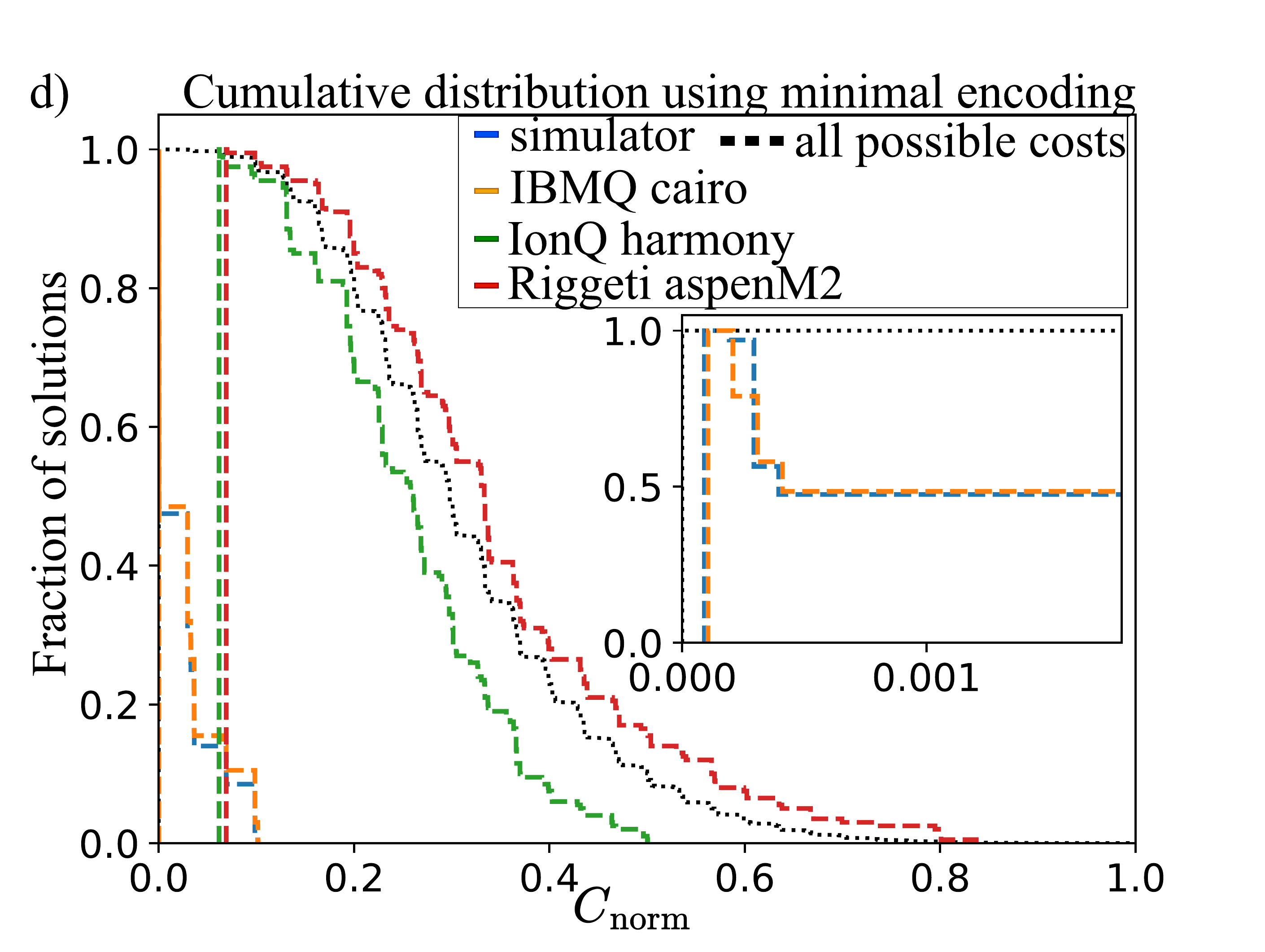} 
\end{tabular}
    
    \caption{Comparison between the full encoding and minimal encoding for a $n_c=16$ route problem, using $n_q=16$ qubits for the full encoding and $n_q=5$ qubits for the minimal encoding. 
    a, c) Optimization runs over 20 starting points. Shaded region shows minimum and maximum value of the 20 runs obtained at each optimization iteration.
    b, d) Cumulative distribution of bitstrings obtained using $n_{\textrm{shots}} = 10000$ shots on a simulator (blue) , IBMQ (orange), IonQ (green), Riggeti (red) quantum backends compared with the distribution of all possible cost function values (black).
    Inset shows zoomed in versions of the cumulative distributions close to $C_{\textrm{norm}} = 0$.
    }
\label{fig:Results16}

\end{figure*}

\subsection{Quantum assisted hybrid algorithms}

Quantum assisted solvers \cite{bharti2020quantum, bharti2021iterative} were recently introduced as a means of solving ground state problems to Hamiltonians of the form $H = \sum_{i=1}^n \beta_i U_i$ without using a classical-quantum feedback loop. A set of efficiently implementable unitaries $V_j$ is chosen to prepare states $\ket{\phi_j} = V_j \ket{0}$ used in constructing an ansatz of the form

\begin{equation}
\ket{\psi (\vec{\alpha})} = \sum_{j=1}^m \alpha_j \ket{\phi_j}=\sum_{j=1}^m \alpha_j V_j \ket{0}
 \label{Eq:QAEAnsatz}
\end{equation} 

where $\vec{\alpha}$ is a vector of complex values.
Once the ansatz is  constructed, a quantum computer is used to calculate the overlap matrices $D_{jk} = \sum_i^n \beta_i \langle \phi_j | U_i |\phi_k \rangle $ and $E_{jk} = \langle \phi_j|\phi_k \rangle$.

To find the optimal values of $\vec{\alpha}$, the following quadratic optimization problem can be solved classically,

\begin{align}
    \textrm{minimize } & \vec{\alpha}^\top {D} \vec{\alpha} \notag \\
    \textrm{subject to } & \vec{\alpha}^\top {E} \vec{\alpha} = 1 \notag
\label{Eq:QAEopt}
\end{align}

where \({D}\), \({E}\) are matrices constructed using \(D_{jk}\), \(E_{jk}\) and written in compact form.

The methods discussed above require mapping a qubit to each binary variable. 
This requirement limits the use of traditional variational approaches when attempting to solve an industry scale problem on NISQ hardware. 
In the following section, we will outline how the encoding scheme in \cite{Tan_2021} can be used to reduce the number of qubits required for a hardware efficient variational approach which can then be applied to the VRPTW problem.

\section{Systematic Encoding of Binary Variables}

The encoding scheme in \cite{Tan_2021} starts off by dividing the $n_c$ classical variables within the QUBO problem into subgroups of $n_a$ variables. 
$n_a$ qubits are then used to represent the values of the binary variables within those subgroups, and the basis states of $n_r = \log_2 (\frac{n_c}{n_a})$ register qubits are used as addresses to denote which subgroup should take on the values that the ancilla qubits represent. 
The smallest possible subgroup allowed consists of only one classical variable per subgroup where $n_a = 1$. We refer to this grouping as the {\it minimal encoding}. 
The largest possible subgroup allowed is when the subgroup consists of all the classical variables, i.e. $n_a = n_c$, in which case only one possible subgroup exists and no register qubits are needed i.e. $n_q = n_a = n_c$. We refer to this as the {\it full encoding} and is the one-to-one qubit to variable mapping typically used in the standard approaches described above.

\subsection{Minimal Encoding}
\label{sec:minimalencoding}

The parametrized quantum state can be expressed as
\begin{equation}
    \ket{\psi_{\textrm{me}} ({\vec \theta})} = \sum_{k=1}^{n_c} \beta_k(\vec\theta) [a_k(\vec\theta) \ket{0}_a + b_k(\vec\theta) \ket{1}_a]\otimes\ket{\phi_k}_r \label{Eq:MinEncoding} 
\end{equation}

where \{$\ket{0}_a$,$\ket{1}_a$\} refer to the quantum state of the $n_a=1$ qubit and \{$\ket{\phi_k}_r$\} refer to the quantum state of the register qubits.
The cost function for the minimal encoding is the same as \eqref{Eq:QUBOprob}. However, the probability $P(\vec{x_i})$ of obtaining a particular bitstring $\vec{x_i}$ is now calculated differently using the state given in \eqref{Eq:MinEncoding}. 

From this state, we can construct a probability distribution using the coefficients to obtain the probability of sampling a bitstring $\vec{x}$ from the state. This can be done by letting the probability of the $k^{\rm th}$ binary variable in $\vec{x}$ being $x_k = 0$ and $x_k = 1$ be $P(x_k = 0) = |a_k(\vec\theta)|^2$ and $P(x_k = 1) = |b_k(\vec\theta)|^2$ respectively. This gives the total probability $P(\vec{x_i})$ of obtaining a particular bitstring $\vec{x_i}$ as 
\begin{align}
    P(\vec{x_i})_{\vec{\theta}} = & \prod_{k=1}^{n_c} |b_k (\vec{\theta})|^2\\
    = & \prod_{k=1}^{n_c} P(x_k) _{\vec{\theta}}
 \label{Eq:MinEncodingProb}
\end{align}
for a given set of $\vec{\theta}$. To obtain a classical bitstring post optimization, coefficients of the optimal quantum state $\ket{\psi_{\textrm{me}}(\vec{\theta}_{\textrm{opt}})}$ are estimated using $n_{\textrm{shots}}$ number of shots to obtain the $b_i(\vec{\theta}_{\textrm{opt}})$ coefficients in Eq.~\ref{Eq:MinEncoding}, from which we can construct the probability distribution above. Multiple classical solutions $\vec{x}$ can then be sampled from this probability distribution by assigning the $k^{\textrm{th}}$ bit in each classical solution to be $x_k = 1$ with a probability of $P(x_k = 1) = |b_k|^2$ and $x_k = 0$ with probability $P(x_k = 1) = 1 - |b_k|^2 = |a_k|^2$. 

The minimal encoding allows us to reduce the number of qubits required for a problem with $n_c$ classical variables to $n_q = 1 + \log_2 (n_c)$ qubits, the largest reduction possible using the encoding scheme.

To estimate the cost function for optimization using the encoding scheme, it is possible to use the variational cost function in Eq.~\ref{Eq:QUBOprob} by measuring the output state of the quantum circuit, constructing the probability distribution in Eq.~\ref{Eq:MinEncodingProb} as described above, and then sampling multiple classical solutions. $P(\vec{x}_i)$ in Eq.~\ref{Eq:QUBOprob} is estimated by calculating the fraction of each unique sample $\vec{x}_i$ obtained, over all samples. However, this results in a slight classical overhead at each optimization step as the quantum state is used to construct a probability distribution to sample classical solutions to estimate Eq.~\ref{Eq:QUBOprob}. 

By substituting the probability distribution $\prod_{k=1}^{n_c} P(x_k)$ into $P(\vec{x}_i)$ in Eq.~\ref{Eq:QUBOprob}, we can obtain a cost function that directly depends on the coefficients in our minimal encoding quantum state \eqref{Eq:MinEncoding}. When expressed in the form of projectors, this becomes:
\begin{align}
	C_{\textrm{me}} (\vec\theta)
	& = \sum_{k\neq l}^{n_c} A_{kl}  \frac{\langle \hat P_k^{1}\rangle_{\vec\theta}\langle\hat P_l^{1}\rangle_{\vec\theta}}{\langle \hat P_k\rangle_{\vec\theta}\langle\hat P_l\rangle_{\vec\theta}}  +\sum_{k=1}^{n_c} A_{kk} \frac{\langle \hat P_k^{1}\rangle_{\vec\theta}}{\langle \hat P_k\rangle_{\vec\theta}}
	\label{Eq:C1} \\
    & = \sum_{k\neq l}^{n_c} A_{kl} |b_k|_{\vec{\theta}}^2 |b_l|_{\vec{\theta}}^2 + \sum_{k=1}^{n_c} A_{kk} |b_k|_{\vec{\theta}}^2
\end{align}
which is the minimal encoding cost function as described in \cite{Tan_2021}. The projectors $\hat P_k = \ket{\phi_k}\bra{\phi_k}_r$ are projectors over the register basis states $\ket{\phi_k}_r$ and $\hat P^1_k = \ket{1}\bra{1}_a \otimes \hat P_k$ are projectors over the states where the ancilla is in the $\ket{1}_a$ state. This removes the need to reconstruct classical bitstrings $\vec{x}_i$ to evaluate multiple values of $\vec{x_i}^{\top} \mathcal{A} \vec{x_i}$ at each optimization step. 




This minimal encoding cost function is a sum over $n_c^2$ number of terms, compared to regular encoding approaches that minimize $\expval{\hat{H}_{\textrm{Ising}}}$. Evaluating $\expval{\hat{H}_{\textrm{Ising}}}$ requires summing over a maximum of $2^{n_c}$ terms in the limit of $n_{\textrm{shots}} \rightarrow \infty$, or at most $n_{\textrm{shots}}$ number of terms if $n_{\textrm{shots}} < 2^{n_c}$ shots are used.

For small number of shots used to estimate \eqref{Eq:C1}, some of the register states may not be measured, leading to $\langle \hat P_k\rangle_{\vec\theta} = 0$ in the denominator. Intuitively, this can be interpreted as not having any information on whether to assign the $k^{\textrm{th}}$ bit to be $x_k = 0$ or $x_k = 1$ (with their respective probabilities). In such cases, we manually set $\frac{\langle \hat P_k^{1}\rangle_{\vec\theta}}{\langle \hat P_k\rangle_{\vec\theta}} = 0.5$. This is interpreted as randomly guessing the value of the $k^{\textrm{th}}$ bit to be $0$ or $1$ with a $50\%$ probability.

The minimal encoding, while being able to exponentially reduce the number of qubits required to find solutions to a QUBO problem, comes at a cost of being able to capture classical correlations between the binary variables during sampling. This can be seen from Eq.~\ref{Eq:MinEncodingProb} which, for fixed $\vec{\theta}$ describes a probability distribution of multiple independent variables where the outcome of sampling the $k^{\textrm{th}}$ bit is independent of the outcome of sampling another bit. 
During optimization however, these marginal probabilities depend on the same set of variational parameters, $\vec{\theta}$, when using the ansatz in Fig.~\ref{fig:circuitansatz}. Adjusting $\vec{\theta}$ to change $P(x_k=1)$ for the $k^{\textrm{th}}$ bit may affect the value $P(x_l=1)$ of the $l^{\textrm{th}}$ bit. Regardless, unless the optimal state is able to produce $P(x_k=1)=0$ or $P(x_k=1)=1$ exactly, it is difficult for the minimal encoding state to produce a sharply peaked distribution around the optimal distribution. 

Another trade-off of the minimal encoding is the number of shots needed to properly characterize a classical solution. In traditional mappings of $1$ qubit to a binary variable, each measurement of the quantum circuit yields a single solution $\vec{x}$. For the minimal encoding, multiple shots are needed to properly characterize the $|b_k|^2$ coefficients in the optimal quantum state. For large problem sizes, this can result in large number of shots, especially if the corresponding register probabilities, $|\beta_k|^2$ are very small. Failure to measure these probabilities results in the issues mentioned above, where the values of certain bits have to be guessed, resulting in poorer solutions.

\subsection{Full Encoding}

Problems solved using the full encoding scheme will have each classical variable mapped to its own qubit. The full encoded quantum state is given in (\ref{Eq:FullEncoding}) and the cost function is calculated according to (\ref{Eq:QUBOprob}).
Each measurement of the circuit results in a basis state that represents a specific bitstring $x_i$, whose cost can be calculated according to $C_{\textrm{fe}} = \vec{x_i}^T \mathcal{A} \vec{x_i}$.

\section{Experimental Procedure}
\label{sec:ExperimentalProcedure}
In this paper, we solve VRPTW instances of $N=3, 13, 22, 62$ destinations that includes $n_c = 11, 16, 128$ and $3964$ routes respectively. In the smaller $2$ instances, $2$ methods of encoding are used - the traditional method of having one variable to one qubit, and the minimal encoding requiring $1+ \log_2(n_c)$ qubits. 
For problem sizes $n_c = 128$ and $3964$, only the minimal encoding was used.

\subsection{Instance generation}
\label{sec:instancegeneration}
A particular VRPTW instance is specified with a set of all valid routes $\mathcal{R}$. 
For fully connected problems with \(N\) nodes, finding the optimal route becomes intractable very quickly, as the number of possible routes is given by $R_{max}=\sum^{N}_{i=1}\frac{N!}{(N-i)!}$. 
In real-world problems, nodes may not always be fully connected and \(R \ll R_{max}\). 
All instances considered here were generated from a set $\mathcal{R}$ obtained using a greedy heuristic provided in \cite{Exxon}.

From this point, we need to convert our set of routes into a QUBO problem of the form in \eqref{Eq:QUBO}. Starting from the cost function of the classical optimization problem in \eqref{Eq:VRPCostFn}, and adding additional penalty terms to represent our constraints, we can arrive at a new, unconstrained cost function:
\begin{equation}
    \underset{\vec{x}}{\textrm{min}}\; \sum_r^R c_r x_r + \rho \sum_i^N \left(\sum_r^R \delta_{ir} x_r - 1 \right)^2
    \label{Eq:CostFnConstrained}
\end{equation} 
where $\rho$ is a positive penalty coefficient that serves to increase our cost function by an additional amount whenever a constraint is violated ie. $\sum_r^R \delta_{ir} x_r \neq 1$. Since, in the worst case scenario, we require this penalty to outweigh the maximum total cost, we set $\rho\geq\sum_r^R |c_r|$ \cite{Exxon}. We can expand the squared term and write \eqref{Eq:CostFnConstrained} in the form:
\begin{align}
    \underset{\vec{x}}{\textrm{min}} \; & \sum_r^R c_r x_r \label{Eq:CostFnUnconstrainedExpanded}\\
    &+\rho 
    \sum_i^N
    \left[ 
    \left(\sum_{r}^R \delta_{ir} x_r \right) 
    \left(\sum_{r'}^R \delta_{ir'} x_{r'} \right) \right. \nonumber \\
    &-2 \left. \sum_r^R \delta_{ir} x_r +1\right] \nonumber \\
    = \underset{\vec{x}}{\textrm{min}} \; & \sum_r^R c_r x_r+ \rho
    \sum_i^N
    \sum_r^R \sum_{r'}^R \delta_{ri}^{\top}x_r x_{r'}  \delta_{ir'}  \label{Eq:CostFnUnconstrainedExpanded2}\\
    & 
    - 2\rho\sum_i^N \left(\sum_r^R \delta_{ir} x_r \right)
    \nonumber\\
    & + \rho N \nonumber
\end{align} 
Equation (\ref{Eq:CostFnUnconstrainedExpanded2})  can be written in matrix and vector form:
\begin{equation}
    \underset{\vec{x}}{\textrm{min}} \; \vec{c}^\top \vec{x} + \rho \vec{x}\delta^\top \delta \vec{x} - 2\rho \left(\vec{1}^\top \delta \right)  \vec{x} + \rho N
    \label{Eq:costfunctionmatrixform}
\end{equation}
from which we can identify the quadratic, linear, and constant terms.
By setting matrix \( \mathcal A=\text{diag}(\vec{c})+\rho \delta^\top\delta-\text{diag}(2 \rho \vec{1}^\top \delta)\), where $\text{diag}(\vec{c})$ denotes a matrix with $\vec{c}$ in its diagonal, we arrive at the definition of a QUBO problem in \eqref{Eq:QUBO}.




\subsection{Optimization procedure}
\label{sec:VI.B}

Each problem is set up with a basic variational ansatz with $L=4$ layers of the gates exemplified in Fig. \ref{fig:circuitansatz}.
We used ADAM \cite{zhang2018improved}, a gradient based classical optimizer, to find the optimal parameters for our circuit. 
During each iteration, the parameter shift rule \cite{crooks2019gradients} was used to calculate the gradient of each parameter.
The final parameters were used to run optimal circuits using a noise-free simulator in Pennylane \cite{bergholm2022pennylane} and, depending on the problem sizes, on the cairo or guadalupe quantum backends from IBMQ \cite{ibmq}, the Aspen-M2 quantum backend from Rigetti via the AWS Braket \cite{braket}, and on the Harmony and Aria-1 quantum backends from IonQ \cite{ionq}.
In all problem instances for both the minimal and full encoding case, $20$ randomly seeded starting points, $\vec{\theta}_{\textrm{init}}$, were used.
$10$ classical solutions were sampled from the post-optimization quantum state for a total of $200$ classical solutions per problem instance, and compared according to a cumulative distribution of their normalized cost function value:
\begin{equation}
	C_{\textrm{norm}}=\frac{(C(\vec{\theta})-C_{\textrm{min}})}{(C_{\textrm{max}}-C_{\textrm{min}})}
 \label{Eq:normalizedcf}
\end{equation}
where $C_{\textrm{min}}$ and $C_{\textrm{max}}$ are the minimum and maximum cost function values associated with the problem instances, found using Gurobi \cite{Gurobi}.
Classical solutions were drawn from the minimal encoding quantum state as described in section \eqref{sec:minimalencoding}.
In the case of full encoded quantum states, to obtain the probability distribution of the solutions we simply measure the quantum circuit in the computational basis. By doing so, we are able to compare these two encoding techniques based on the classical solutions obtained from each method and their associated costs for each VRPTW instance.



\begin{figure}
    
    \includegraphics[width=\columnwidth]{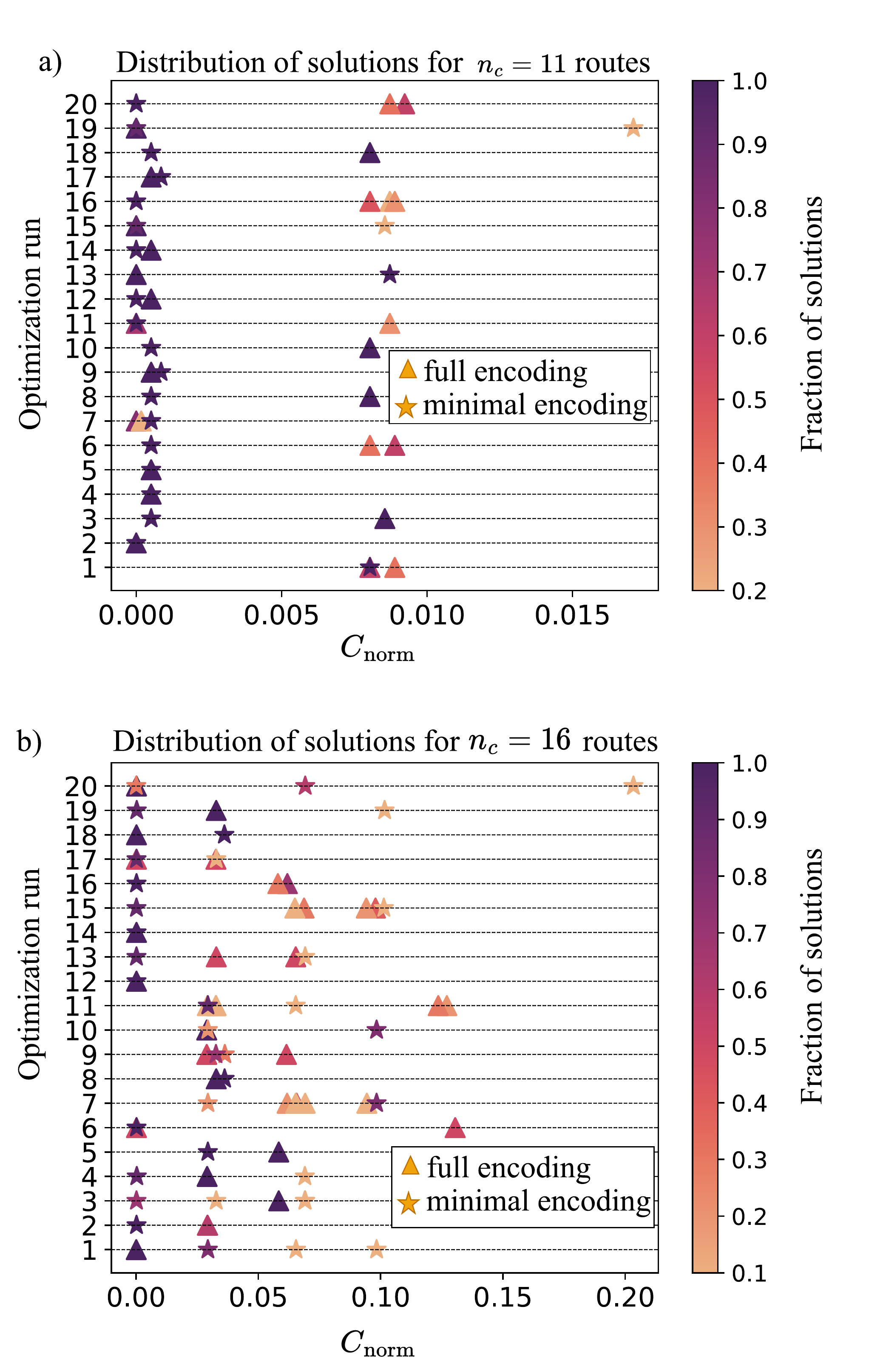}
    \caption{Distribution of solutions obtained using the minimal (stars) and full encoding (triangles) for a) $n_c= 11$ and b) $n_c = 16$ for $20$ different optimization runs from the IBMQ backends. 
For each optimization run, the final quantum state is prepared on the IBMQ backends according to the optimized parameters, and $10$ classical solutions are sampled according to (\ref{Eq:FullEncoding}) for the full encoding and (\ref{Eq:MinEncodingProb}) for the minimal encoding. In both the $11$R and $16$R instances, the minimal encoding is able to produce classical solutions of similar quality while maintaining a tighter spread in terms of cost function values.}
    \label{fig:fig_4}
\end{figure}

\section{Comparison of minimal encoding with full encoding scheme}

The optimization run and approximate solutions obtained for the smaller problem sizes of $n_c = 11, 16$ can be seen in Fig. \ref{fig:Results11} and Fig. \ref{fig:Results16} respectively.
Optimization was done using different number of shots, $n_{\textrm{shots}}$, depending on the size of the problem, and compared with runs completed with full access to the statevector (i.e. $n_{\textrm{shots}} \rightarrow \infty$).
For such small problem sizes, the number of shots required to converge to a solution similar to that obtained using statevector simulation is well within the capabilities of current quantum devices.
With sufficient shots to estimate the quantum state, ADAM is able to find solutions of similar quality for both the minimal encoding and the full encoding, despite the minimal encoding requiring fewer qubits.

\begin{figure*}
\centering

\begin{tabular}{ccc}
\includegraphics[width=0.33\linewidth]{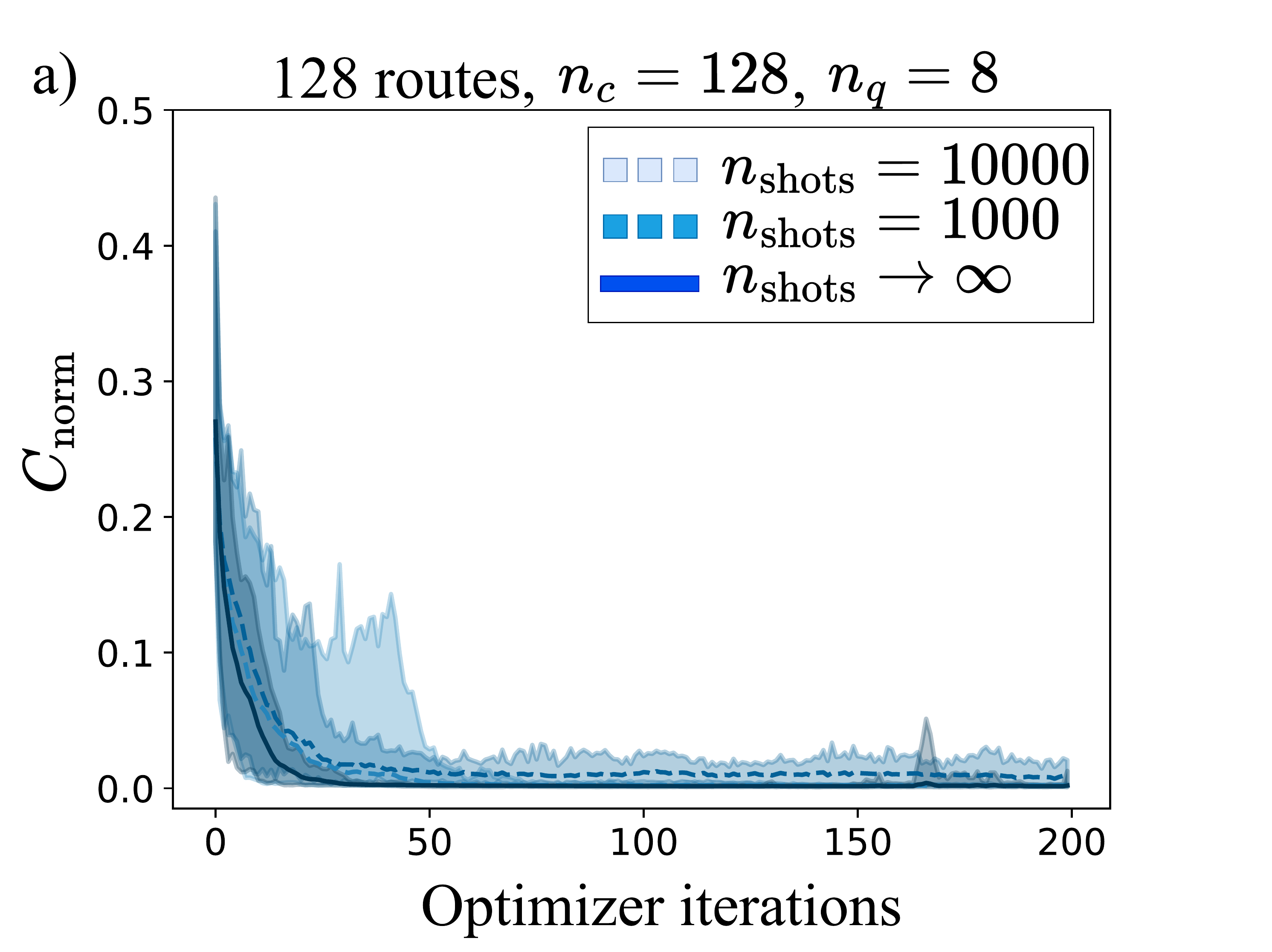} &
\includegraphics[width=0.33\linewidth]{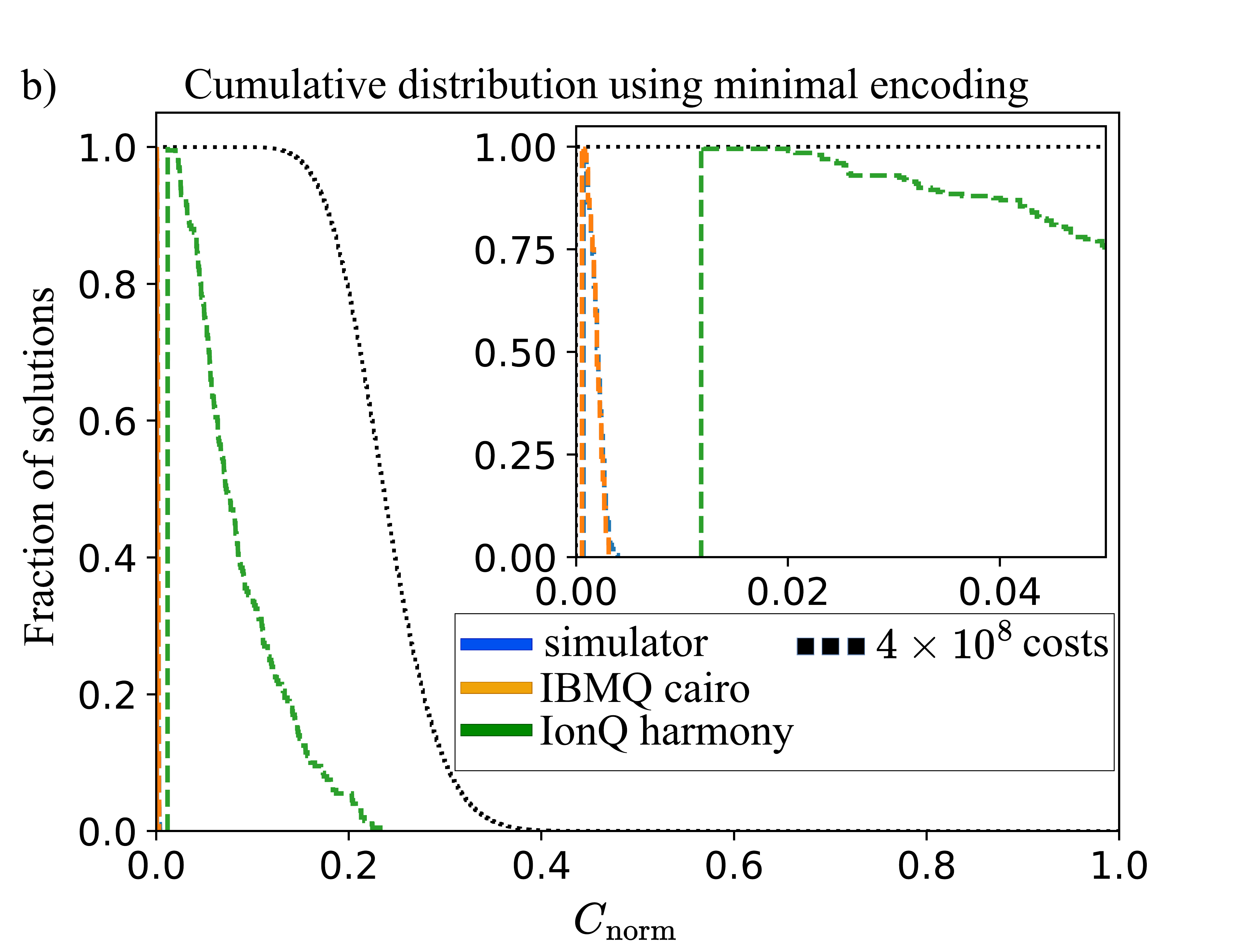} &
\includegraphics[width=0.33\linewidth]{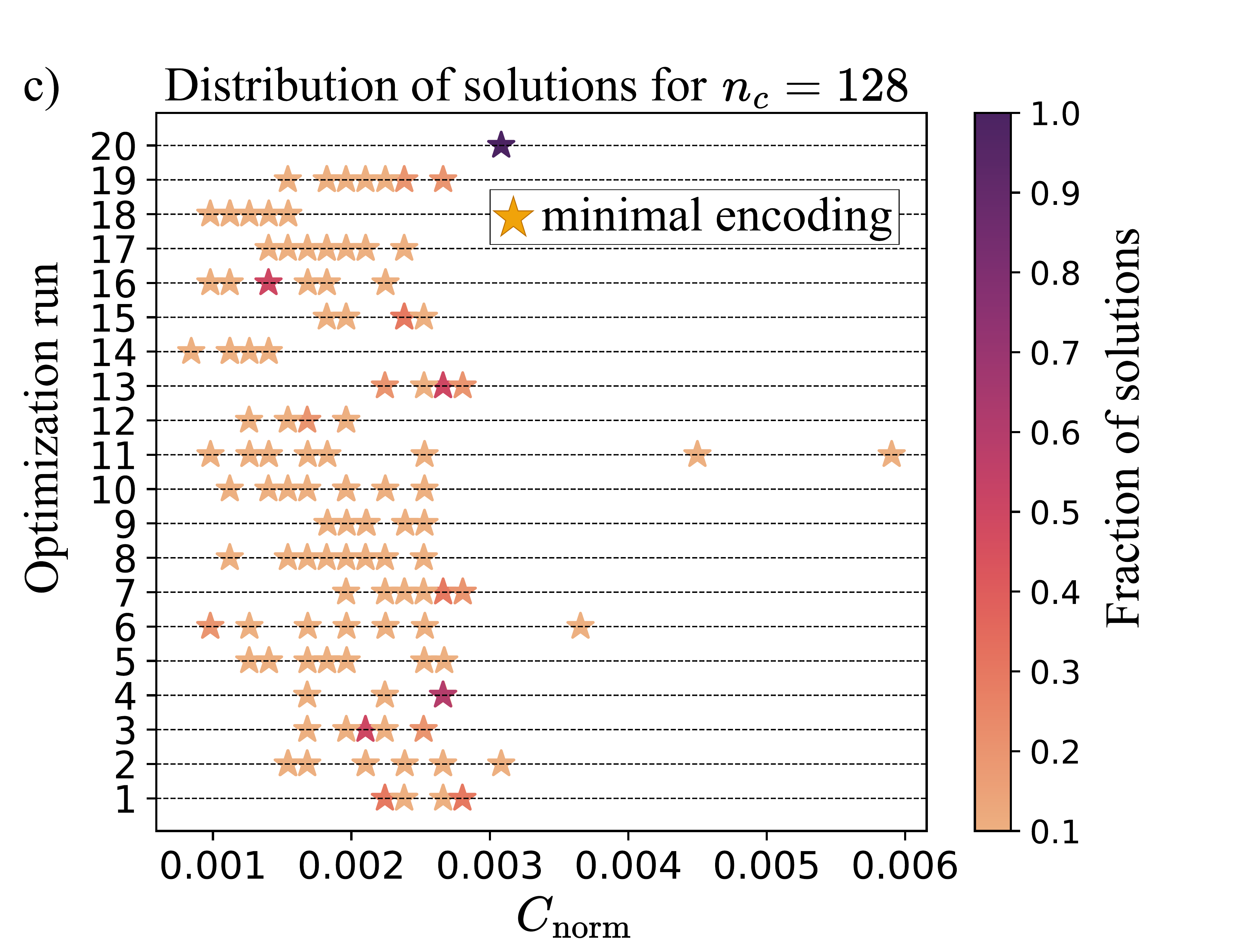}\\
\includegraphics[width=0.33\linewidth]{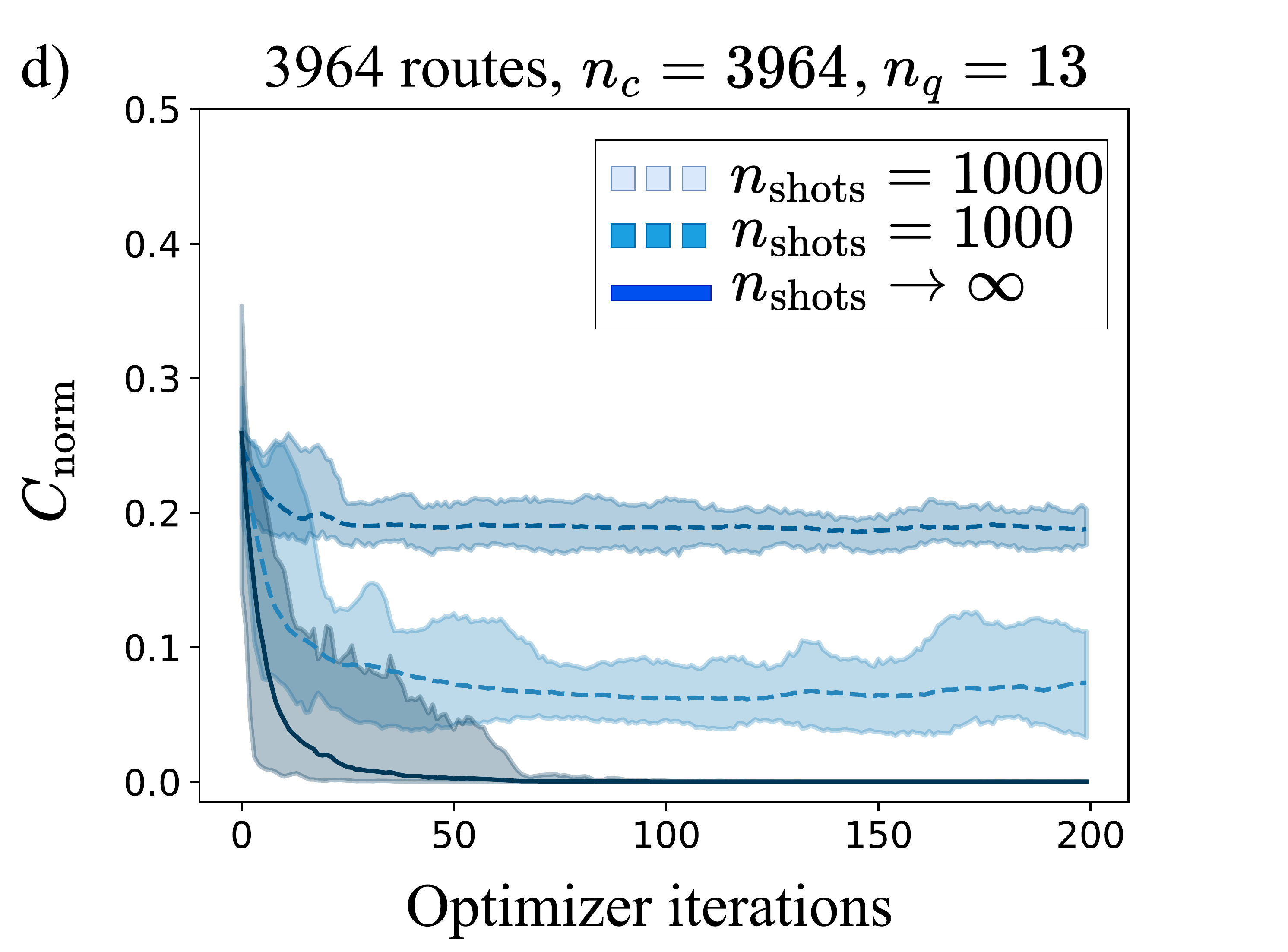} &
\includegraphics[width=0.33\linewidth]{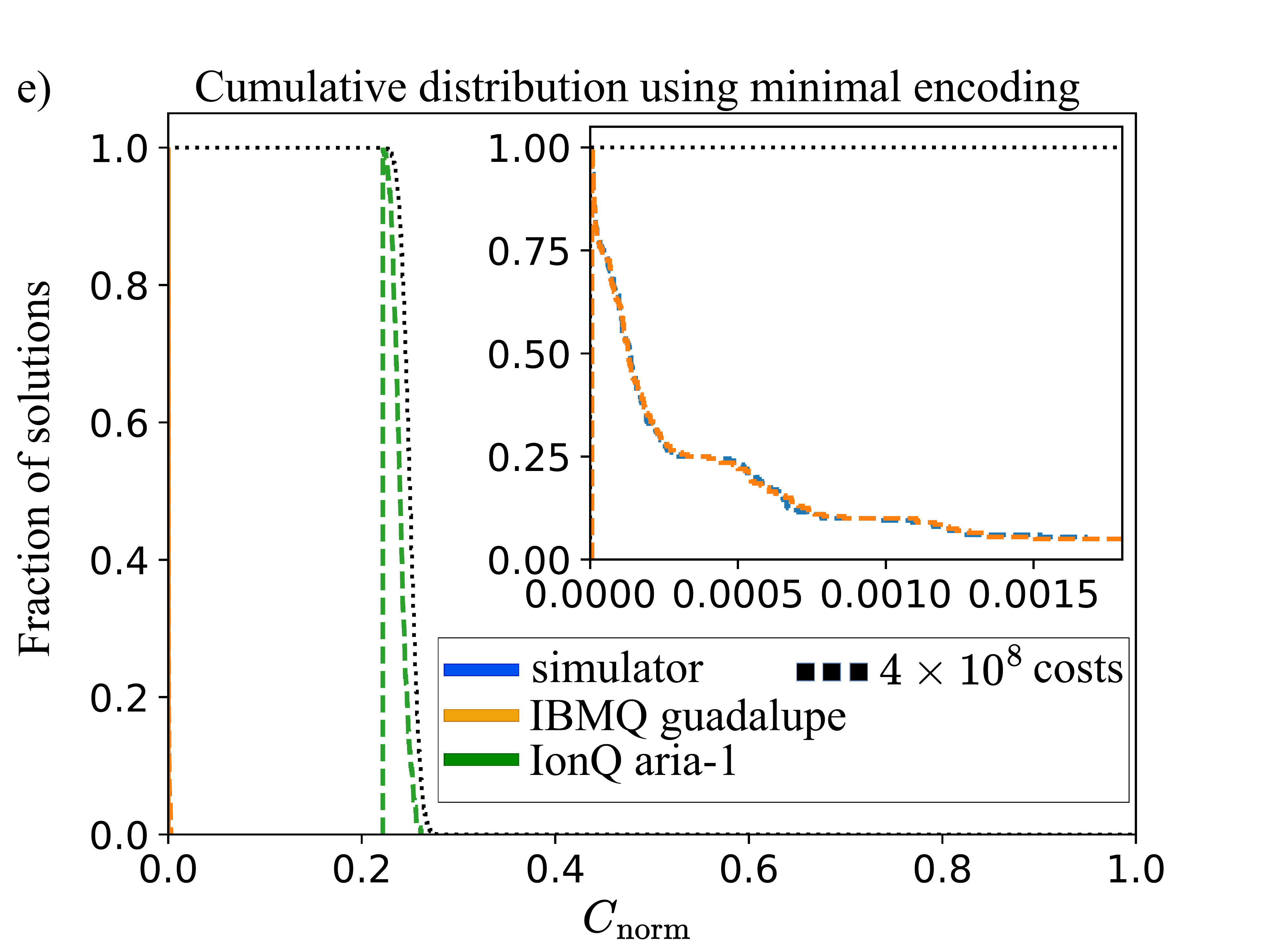} &
\includegraphics[width=0.33\linewidth]{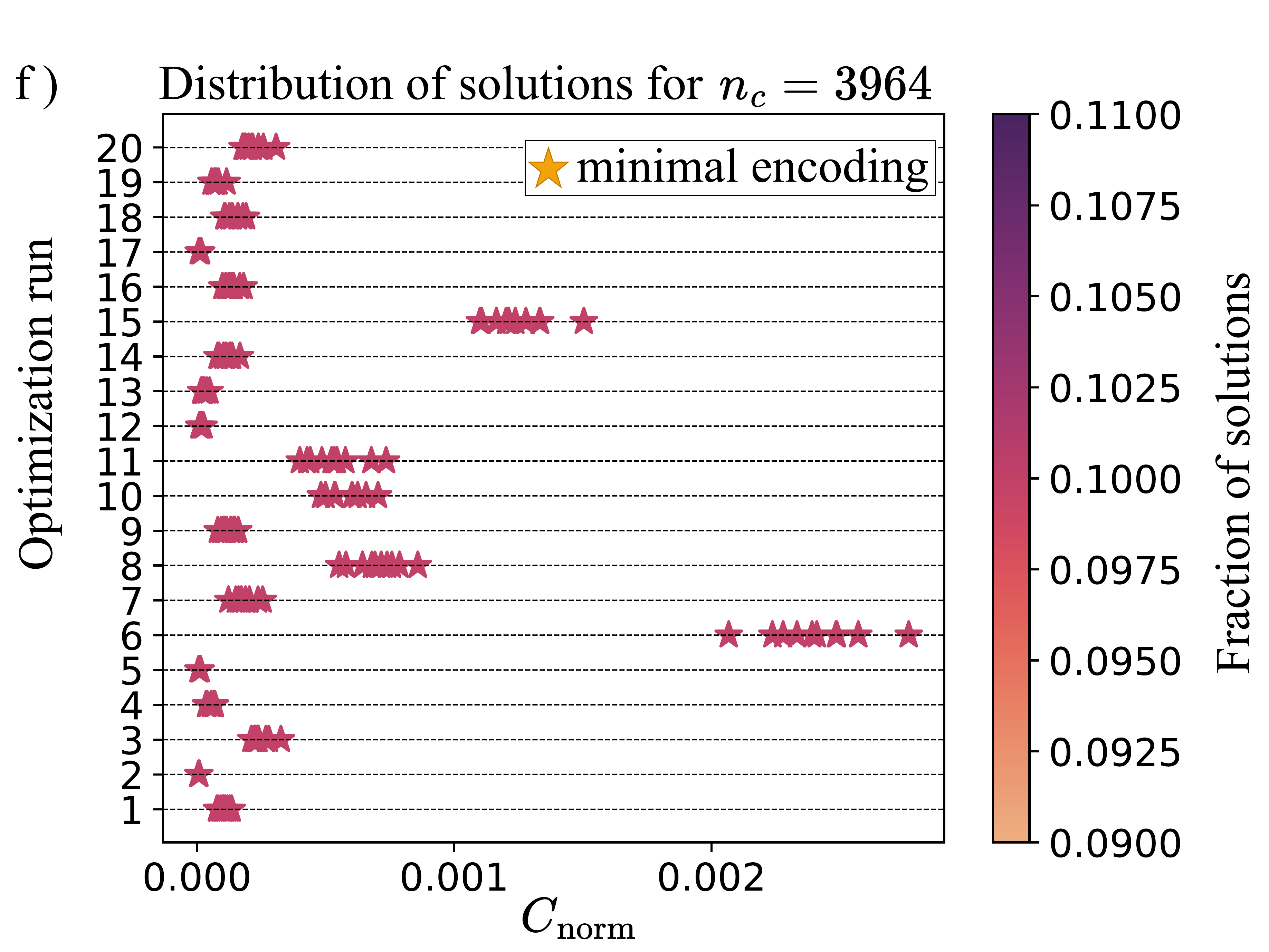}
\end{tabular}
    
    \caption{Optimization runs, cumulative distribution and distribution of solutions using minimal encoding for the $n_c=128$ instance using $n_q = 8$ qubits and $n_c=3964$ instance using $n_q = 13$ qubits. a,d) Optimization runs over 20 starting points. b,e) Cumulative distribution of solutions obtained using minimal encoding. Black dotted curves represents the cumulative distribution of $4\times 10^8$ randomly generated solutions. Blue, orange, and green curves display the cumulative distribution of solutions obtained using minimal encoding with $n_{\textrm{meas}}=10000$ shots on a noise free simulation and on IBMQ, IonQ quantum backends respectively. c,f) Distribution of solutions obtained from the minimal encoding state used to find approximate solutions to the $n_c = 128$R and $n_c = 3964$R problem instances. The final parameters from each of the $20$ starting points are used to produce the final quantum state using the devices offered by IBMQ, and $10$ classical solutions were sampled from each state according to (\ref{Eq:MinEncodingProb}).
    }
\label{fig:fig_5}

\end{figure*}

In both the minimal encoding and full encoding case, classical solutions obtained from quantum states produced by the IBMQ quantum backends consistently outperformed those obtained from the IonQ and Rigetti backends.
Fig. \ref{fig:Results11} and Fig. \ref{fig:Results16} also shows the IonQ devices being able to produce better quality solutions for the full encoded cases compared to solutions sampled from the minimal encoding.
However, due to scheduling issues, problem instances requiring fewer qubits (such as those using the minimal encoding) were executed on harmony instead of Aria-1 qpu, and further testing is required to conclude the source of these differences for the IonQ backends. 

Fig. \ref{fig:fig_4} compare the distribution of classical solutions obtained from the minimal encoding and full encoding for both the $11$R and $16$R instances using the quantum devices from IBMQ, due to their close performance to simulated results.

In both the $11$R and $16$R instances, a good portion of the classical solutions sampled from both minimal encoding and full encoding states are able to produce solutions of similar quality for all the optimization runs. 
However the minimal encoding state is able to produce a tighter clustering of solutions sampled compared to runs using the full encoding.
For non-degenerate problem instances such as the VRPTW instances considered here, this can be explained by having the minimal encoding state converging to a state that produces a high probability of sampling a \textit{single} low-cost solution, and sampling classical solutions from this final state according to the probabilities in (\ref{Eq:MinEncodingProb}), this yields this good solution with high probability. 
However, being unable to capture any classical correlations means that solutions with poor cost function values can also be sampled with non-zero probability. 

This is in contrast to the full encoding case where a good cost function value suggests that the maximization of many good solutions with low cost, ideally peaking at the optimal solution. 
Therefore, sampling from such a post-optimized state using the full encoding is likely to produce other nearby solutions in terms of their cost function values.


\section{Finding solutions to VRPTW problems of larger problem sizes}

To further push the boundaries on the sizes of industry relevant problems that can be solved on quantum devices, we apply the minimal encoding method to find approximate solutions to VRPTW problems of $n_c=128$ and $n_c=3964$ classical variables, using the methods outlined in section \ref{sec:VI.B}. This requires $n_q = \log (128) + 1 = 8$ and $n_q = \lceil \log (3964)  \rceil + 1 = 13$ qubits respectively, where $\lceil \cdot  \rceil $ is the ceiling function. 
While the $128$ route instance in Fig. \ref{fig:fig_5} (a-c) was able to find decent solutions using similar number of shots as the $11$ and $16$ route instances, the number of resources required to obtain decent solutions increases significantly for the $3964$ route instance, as seen in Fig. \ref{fig:fig_5}d where the number of shots limits the quality of solutions that can be obtained. 
Solutions obtained using the IBMQ quantum backends, for both larger instances, fall within the range of those obtained in noise free simulations, as observed in Figures \ref{fig:fig_5}b,e.  Although as seen in the inset axes of the aforementioned figures, the exact optimal solution was not found for any of the two bigger problem instances. 


Similar to the results for the smaller problem instances in Fig. \ref{fig:Results11} and Fig. \ref{fig:Results16}, the solutions obtained from the IBMQ backends were far superior to those obtained from the IonQ devices using the minimal encoding.

For the $3964$-route instance, the minimal encoding state has $2^{12} = 4096$ register states, with only $3964$ of them being assigned to each of the routes. Having only $n_{\textrm{shots}} = 1000$ means that many of these register states are not measured, resulting in having to manually set $\frac{\langle \hat P_k^{1}\rangle_{\vec\theta}}{\langle \hat P_k\rangle_{\vec\theta}} = 0.5$ for many of the terms in \eqref{Eq:C1}. After expressing each state in the form given in  \eqref{Eq:MinEncoding}, these leave many register states unmeasured, resulting in having to guess for multiple routes whether they should be included in the solution or not, leading to poor quality solutions overall.
Increasing the number of shots to $n_{\textrm{shots}} = 10,000$, we observe a lower cost function value, as more register states are able to be measured. However the quantum state has $2^{13} = 8192$ basis vectors in the computational basis, and $10,000$ shots is still insufficient to accurately characterize all coefficients associated with the basis vectors. For large problem instances such as these, it is expected that more shots would be required, although this was not explored in this work due to cost limitations.

As the problem size increases, we observe from Fig. \ref{fig:fig_5}c that the classical solutions sampled from the optimized minimal encoding state for the $128$R instance tend to be more spread out rather than converging on a state which provides a high probability of sampling a single solution, suggesting suboptimal convergence. 
However, the final state is still able to produce bitstrings that have low cost function values.

For the $3964$R instance in Fig. \ref{fig:fig_5}f, we observe a tighter clustering of solutions obtained, although we see several initial starting points that did not converge favourably.

\section{Conclusion}

In this work, we applied the encoding scheme described in \cite{Tan_2021} to a common optimization problem faced within the industry, namely the vehicle routing optimization problem with time windows. 
We compared the results obtained from optimization runs using the minimal encoding and the full encoding for problem sizes of 11 routes and 16 routes and classical solutions sampled from the final states prepared on quantum backends offered by IBMQ, IonQ, and Rigetti via the AWS Braket.
We went beyond the usual boundaries to problem sizes that can be solved using quantum devices by testing the minimal encoding on problem sizes involving 128 and 3964 routes, and compared the solutions obtained using quantum states prepared by the IBMQ and IonQ quantum devices.

The formulation of the VRPTW as a QUBO results in a cost function spectrum where feasible solutions lie close together and are separated from infeasible solutions by a large gap.
This spectrum can make it more difficult for feasible solutions to be found using the minimal encoding due to the quantum state not being able to capture any classical correlations between the variables, leading to increased variance in the cost function during optimization, and due to the quality of solutions becoming more sensitive to the optimization parameters (e.g. the number of shots). 

Our results show that optimization runs using the minimal encoding were still able to find classical solutions of similar quality to runs using the full encoding scheme, despite using much fewer qubits and the limitation of not being able to capture any classical correlations.
Future work in this direction to improve performance could include optimizing over the log of the QUBO matrices to reduce variances within the cost function landscape, using additional qubits to capture (anti-)correlations between multiple routes that contain the same destinations as these routes should not be part of the same solution due to the constraints, or to use additional post-processing methods to obtain feasible solutions (solutions which satisfy all the constraints) from infeasible ones,  
This should allow for approximate valid solutions to still be found despite some of the optimization runs terminating in a local minima where only some of the constraints are satisfied.





{\it Acknowledgments.}--- The authors would like to thank Dimitar Trenev and Stuart Harwood from ExxonMobil for their support on formulating the VRPTW instances. This research is supported by the National Research Foundation, Prime Minister’s Office, Singapore under its Quantum Engineering Programme (NRF2021-QEP2-02-P02) and EU HORIZON-Project101080085—QCFD.

\bibliographystyle{unsrt}
\bibliography{VRP_bibliography}

\begin{thebibliography}{100}

\bibitem{vikstaal2020applying}
Pontus Vikst{\aa}l, Mattias Gr{\"o}nkvist, Marika Svensson, Martin Andersson,
  G{\"o}ran Johansson, and Giulia Ferrini.
\newblock Applying the quantum approximate optimization algorithm to the
  tail-assignment problem.
\newblock {\em Physical Review Applied}, 14(3):034009, 2020.

\bibitem{Exxon}
Stuart Harwood, Claudio Gambella, Dimitar Trenev, Andrea Simonetto, David
  Bernal, and Donny Greenberg.
\newblock Formulating and solving routing problems on quantum computers.
\newblock {\em IEEE Transactions on Quantum Engineering}, 2:1--17, 2021.

\bibitem{kochenberger2014}
G.~Kochenberger, J.K. Hao, F.~Glover, M.~Lewis, Z.P L{\"u}, H.B Wang, and
  Y.~Wang.
\newblock The unconstrained binary quadratic programming problem: a survey.
\newblock {\em Journal of Combinatorial Optimization}, 28(1):58--81, 2014.

\bibitem{glover2018tutorial}
Fred Glover, Gary Kochenberger, and Yu~Du.
\newblock A tutorial on formulating and using qubo models.
\newblock {\em arXiv preprint arXiv:1811.11538}, 2018.

\bibitem{salehi2022unconstrained}
{\"O}zlem Salehi, Adam Glos, and Jaros{\l}aw~Adam Miszczak.
\newblock Unconstrained binary models of the travelling salesman problem
  variants for quantum optimization.
\newblock {\em Quantum Information Processing}, 21(2):67, 2022.

\bibitem{dominguez2023encoding}
Federico Dominguez, Josua Unger, Matthias Traube, Barry Mant, Christian Ertler,
  and Wolfgang Lechner.
\newblock Encoding-independent optimization problem formulation for quantum
  computing.
\newblock {\em arXiv preprint arXiv:2302.03711}, 2023.

\bibitem{mattesi2023financial}
Mirko Mattesi, Luca Asproni, Christian Mattia, Simone Tufano, Giacomo Ranieri,
  Davide Caputo, and Davide Corbelletto.
\newblock Financial portfolio optimization: a qubo formulation for sharpe ratio
  maximization, 2023.

\bibitem{Glover2022}
Fred Glover, Gary Kochenberger, Rick Hennig, and Yu~Du.
\newblock Quantum bridge analytics i: a tutorial on formulating and using qubo
  models.
\newblock {\em Annals of Operations Research}, 314(1):141--183, Jul 2022.

\bibitem{Streif2021}
Michael Streif, Sheir Yarkoni, Andrea Skolik, Florian Neukart, and Martin Leib.
\newblock Beating classical heuristics for the binary paint shop problem with
  the quantum approximate optimization algorithm.
\newblock {\em Phys. Rev. A}, 104:012403, Jul 2021.

\bibitem{venturelli2015quantum}
Davide Venturelli, Dominic~JJ Marchand, and Galo Rojo.
\newblock Quantum annealing implementation of job-shop scheduling.
\newblock {\em arXiv preprint arXiv:1506.08479}, 2015.

\bibitem{adelomou2020using}
Atchade~Parfait Adelomou, Elisabet~Golobardes Ribe, and Xavier~Vilasis Cardona.
\newblock Using the parameterized quantum circuit combined with
  variational-quantum-eigensolver (vqe) to create an intelligent social
  workers' schedule problem solver, 2020.

\bibitem{dantzig1959truck}
George~B Dantzig and John~H Ramser.
\newblock The truck dispatching problem.
\newblock {\em Management science}, 6(1):80--91, 1959.

\bibitem{braekers2016vehicle}
Kris Braekers, Katrien Ramaekers, and Inneke Van~Nieuwenhuyse.
\newblock The vehicle routing problem: State of the art classification and
  review.
\newblock {\em Computers \& industrial engineering}, 99:300--313, 2016.

\bibitem{toth2002vehicle}
Paolo Toth and Daniele Vigo.
\newblock {\em The vehicle routing problem}.
\newblock SIAM, 2002.

\bibitem{lin2014survey}
Canhong Lin, King~Lun Choy, George~TS Ho, Sai~Ho Chung, and HY~Lam.
\newblock Survey of green vehicle routing problem: past and future trends.
\newblock {\em Expert systems with applications}, 41(4):1118--1138, 2014.

\bibitem{moghdani2021green}
Reza Moghdani, Khodakaram Salimifard, Emrah Demir, and Abdelkader Benyettou.
\newblock The green vehicle routing problem: A systematic literature review.
\newblock {\em Journal of Cleaner Production}, 279:123691, 2021.

\bibitem{kara2007energy}
Imdat Kara, Bahar~Y Kara, and M~Kadri Yetis.
\newblock Energy minimizing vehicle routing problem.
\newblock In {\em Combinatorial Optimization and Applications: First
  International Conference, COCOA 2007, Xi’an, China, August 14-16, 2007.
  Proceedings 1}, pages 62--71. Springer, 2007.

\bibitem{F_Barahona_1982}
F~Barahona.
\newblock On the computational complexity of ising spin glass models.
\newblock {\em Journal of Physics A: Mathematical and General}, 15(10):3241,
  oct 1982.

\bibitem{Bharti2021}
Kishor Bharti, Alba Cervera-Lierta, Thi~Ha Kyaw, Tobias Haug, Sumner
  Alperin-Lea, Abhinav Anand, Matthias Degroote, Hermanni Heimonen, Jakob~S.
  Kottmann, Tim Menke, Wai-Keong Mok, Sukin Sim, Leong-Chuan Kwek, and Al\'an
  Aspuru-Guzik.
\newblock Noisy intermediate-scale quantum algorithms.
\newblock {\em Rev. Mod. Phys.}, 94:015004, Feb 2022.

\bibitem{sawaya2022encoding}
Nicolas~PD Sawaya, Albert~T Schmitz, and Stuart Hadfield.
\newblock Encoding trade-offs and design toolkits in quantum algorithms for
  discrete optimization: coloring, routing, scheduling, and other problems.
\newblock {\em arXiv preprint arXiv:2203.14432}, 2022.

\bibitem{FINNILA1994343}
A.B. Finnila, M.A. Gomez, C.~Sebenik, C.~Stenson, and J.D. Doll.
\newblock Quantum annealing: A new method for minimizing multidimensional
  functions.
\newblock {\em Chemical Physics Letters}, 219(5):343--348, 1994.

\bibitem{Kadowaki1998QA}
Tadashi Kadowaki and Hidetoshi Nishimori.
\newblock Quantum annealing in the transverse ising model.
\newblock {\em Phys. Rev. E}, 58:5355--5363, Nov 1998.

\bibitem{Farhi2014QAOA1}
Edward Farhi, Jeffrey Goldstone, and Sam Gutmann.
\newblock A quantum approximate optimization algorithm.
\newblock {\em arXiv preprint arXiv:1411.4028}, 2014.

\bibitem{Farhi2014QAOA2}
Edward Farhi, Jeffrey Goldstone, and Sam Gutmann.
\newblock A quantum approximate optimization algorithm applied to a bounded
  occurrence constraint problem.
\newblock {\em arXiv preprint arXiv:1412.6062}, 2014.

\bibitem{Farhi2019}
Edward Farhi, Jeffrey Goldstone, Sam Gutmann, and Leo Zhou.
\newblock The quantum approximate optimization algorithm and the
  sherrington-kirkpatrick model at infinite size.
\newblock {\em arXiv preprint arXiv:1910.08187}, 2019.

\bibitem{blekos2023review}
Kostas Blekos, Dean Brand, Andrea Ceschini, Chiao-Hui Chou, Rui-Hao Li, Komal
  Pandya, and Alessandro Summer.
\newblock A review on quantum approximate optimization algorithm and its
  variants, 2023.

\bibitem{hadfield2019quantum}
Stuart Hadfield, Zhihui Wang, Bryan O’gorman, Eleanor~G Rieffel, Davide
  Venturelli, and Rupak Biswas.
\newblock From the quantum approximate optimization algorithm to a quantum
  alternating operator ansatz.
\newblock {\em Algorithms}, 12(2):34, 2019.

\bibitem{Zhou2018}
Leo Zhou, Sheng-Tao Wang, Soonwon Choi, Hannes Pichler, and Mikhail~D. Lukin.
\newblock Quantum approximate optimization algorithm: Performance, mechanism,
  and implementation on near-term devices.
\newblock {\em arXiv preprint arXiv:1812.01041}, 2018.

\bibitem{mohammadbagherpoor2021exploring}
Hamed Mohammadbagherpoor, Patrick Dreher, Mohannad Ibrahim, Young-Hyun Oh,
  James Hall, Richard~E Stone, and Mirela Stojkovic.
\newblock Exploring airline gate-scheduling optimization using quantum
  computers, 2021.

\bibitem{chatterjee2021variational}
Turbasu Chatterjee, Shah~Ishmam Mohtashim, and Akash Kundu.
\newblock On the variational perspectives to the graph isomorphism problem,
  2021.

\bibitem{Pablo2021}
Pablo D\'{\i}ez-Valle, Diego Porras, and Juan~Jos\'e Garc\'{\i}a-Ripoll.
\newblock Quantum variational optimization: The role of entanglement and
  problem hardness.
\newblock {\em Phys. Rev. A}, 104:062426, Dec 2021.

\bibitem{Egger9222275}
Daniel~J. Egger, Claudio Gambella, Jakub Marecek, Scott McFaddin, Martin
  Mevissen, Rudy Raymond, Andrea Simonetto, Stefan Woerner, and Elena Yndurain.
\newblock Quantum computing for finance: State-of-the-art and future prospects.
\newblock {\em IEEE Transactions on Quantum Engineering}, 1:1--24, 2020.

\bibitem{amaro2022filtering}
David Amaro, Carlo Modica, Matthias Rosenkranz, Mattia Fiorentini, Marcello
  Benedetti, and Michael Lubasch.
\newblock Filtering variational quantum algorithms for combinatorial
  optimization.
\newblock {\em Quantum Science and Technology}, 7(1):015021, 2022.

\bibitem{Preskill2018NISQ}
John Preskill.
\newblock Quantum {C}omputing in the {NISQ} era and beyond.
\newblock {\em {Quantum}}, 2:79, August 2018.

\bibitem{leymann2020bitter}
Frank Leymann and Johanna Barzen.
\newblock The bitter truth about gate-based quantum algorithms in the nisq era.
\newblock {\em Quantum Science and Technology}, 5(4):044007, 2020.

\bibitem{lau2022nisq}
Jonathan Wei~Zhong Lau, Kian~Hwee Lim, Harshank Shrotriya, and Leong~Chuan
  Kwek.
\newblock Nisq computing: where are we and where do we go?
\newblock {\em AAPPS Bulletin}, 32(1):27, 2022.

\bibitem{GLOVER2018829}
Fred Glover, Mark Lewis, and Gary Kochenberger.
\newblock Logical and inequality implications for reducing the size and
  difficulty of quadratic unconstrained binary optimization problems.
\newblock {\em European Journal of Operational Research}, 265(3):829--842,
  2018.

\bibitem{Lewis2017}
Mark Lewis and Fred Glover.
\newblock Quadratic unconstrained binary optimization problem preprocessing:
  Theory and empirical analysis.
\newblock {\em Netw.}, 70(2):79–97, September 2017.

\bibitem{Boros2006}
Endre Boros, Peter Hammer, and Gabriel Tavares.
\newblock Preprocessing of unconstrained quadratic binary optimization.
\newblock {\em RUTCOR Research Report}, 01 2006.

\bibitem{Lange2019}
Jan-Hendrik Lange, Bjoern Andres, and Paul Swoboda.
\newblock Combinatorial persistency criteria for multicut and {m}ax-{c}ut.
\newblock In {\em 2019 IEEE/CVF Conference on Computer Vision and Pattern
  Recognition (CVPR)}, pages 6086--6095, 2019.

\bibitem{Ferizovic2020}
Damir Ferizovic, Demian Hespe, Sebastian Lamm, Matthias Mnich, Christian
  Schulz, and Darren Strash.
\newblock {\em Engineering Kernelization for Maximum Cut}, pages 27--41.
\newblock 2020 Proceedings of the Symposium on Algorithm Engineering and
  Experiments (ALENEX), 2020.

\bibitem{Hammer1968}
Peter~L. Hammer and Sergiu Rudeanu.
\newblock {\em Minimization of Nonlinear Pseudo-Boolean Functions}, pages
  113--134.
\newblock Springer Berlin Heidelberg, Berlin, Heidelberg, 1968.

\bibitem{shaydulin2023qaoa}
Ruslan Shaydulin and Marco Pistoia.
\newblock {QAOA} with $n\cdot p\geq 200$, 2023.

\bibitem{Pelofske_2023}
Elijah Pelofske, Andreas Bärtschi, and Stephan Eidenbenz.
\newblock Quantum annealing vs. {QAOA}: 127 qubit higher-order {I}sing problems
  on~{NISQ} computers.
\newblock In {\em Lecture Notes in Computer Science}, pages 240--258. Springer
  Nature Switzerland, 2023.

\bibitem{Lowe_2023}
Angus Lowe, Matija Medvidovi{\'{c}}, Anthony Hayes, Lee~J. O'Riordan, Thomas~R.
  Bromley, Juan~Miguel Arrazola, and Nathan Killoran.
\newblock Fast quantum circuit cutting with randomized measurements.
\newblock {\em {Quantum}}, 7:934, March 2023.

\bibitem{saleem2022divide}
Zain~H. Saleem, Teague Tomesh, Michael~A. Perlin, Pranav Gokhale, and Martin
  Suchara.
\newblock Divide and conquer for combinatorial optimization and distributed
  quantum computation, 2022.

\bibitem{bechtold2023investigating}
Marvin Bechtold, Johanna Barzen, Frank Leymann, Alexander Mandl, Julian Obst,
  Felix Truger, and Benjamin Weder.
\newblock Investigating the effect of circuit cutting in {QAOA} for the
  {M}ax{c}ut problem on {NISQ} devices, 2023.

\bibitem{Peng_2020}
Tianyi Peng, Aram~W. Harrow, Maris Ozols, and Xiaodi Wu.
\newblock Simulating large quantum circuits on a small quantum computer.
\newblock {\em Physical Review Letters}, 125(15), October 2020.

\bibitem{Fujii2022}
Keisuke Fujii, Kaoru Mizuta, Hiroshi Ueda, Kosuke Mitarai, Wataru Mizukami, and
  Yuya~O. Nakagawa.
\newblock Deep variational quantum eigensolver: A divide-and-conquer method for
  solving a larger problem with smaller size quantum computers.
\newblock {\em PRX Quantum}, 3:010346, March 2022.

\bibitem{Mitarai_2021}
Kosuke Mitarai and Keisuke Fujii.
\newblock Constructing a virtual two-qubit gate by sampling single-qubit
  operations.
\newblock {\em New Journal of Physics}, 23(2):023021, February 2021.

\bibitem{Bravyi_2016}
Sergey Bravyi, Graeme Smith, and John~A. Smolin.
\newblock Trading classical and quantum computational resources.
\newblock {\em Physical Review X}, 6(2), June 2016.

\bibitem{Zhang2022}
Yu~Zhang, Lukasz Cincio, Christian F.~A. Negre, Piotr Czarnik, Patrick~J.
  Coles, Petr~M. Anisimov, Susan~M. Mniszewski, Sergei Tretiak, and Pavel~A.
  Dub.
\newblock Variational quantum eigensolver with reduced circuit complexity.
\newblock {\em npj Quantum Information}, 8(1):96, August 2022.

\bibitem{liu2022hybrid}
Chen-Yu Liu and Hsi-Sheng Goan.
\newblock Hybrid gate-based and annealing quantum computing for large-size
  {I}sing problems, 2022.

\bibitem{boost2017partitioning}
M~Boost, SP~Reinhardt, and A~Roy.
\newblock Partitioning optimization problems for hybrid classical/quantum
  execution.
\newblock Technical report, Technical Report, http://www. dwavesys. com, 2017.

\bibitem{tang2022scaleqc}
Wei Tang and Margaret Martonosi.
\newblock Scaleqc: A scalable framework for hybrid computation on quantum and
  classical processors, 2022.

\bibitem{piveteau2023circuit}
Christophe Piveteau and David Sutter.
\newblock Circuit knitting with classical communication, 2023.

\bibitem{glos2020space}
Adam Glos, Aleksandra Krawiec, and Zolt{\'a}n Zimbor{\'a}s.
\newblock Space-efficient binary optimization for variational quantum
  computing.
\newblock {\em npj Quantum Information}, 8(1):39, Apr 2022.

\bibitem{latorre2005image}
Jose~I. Latorre.
\newblock Image compression and entanglement, 2005.

\bibitem{Plewa_Sieńko_Rycerz_2021}
Julia Plewa, Joanna Sieńko, and Katarzyna Rycerz.
\newblock Variational algorithms for workflow scheduling problem in gate-based
  quantum devices.
\newblock {\em COMPUTING AND INFORMATICS}, 40(4):897–929, December 2021.

\bibitem{Tan_2021}
Benjamin Tan, Marc-Antoine Lemonde, Supanut Thanasilp, Jirawat Tangpanitanon,
  and Dimitris~G. Angelakis.
\newblock Qubit-efficient encoding schemes for binary optimisation problems.
\newblock {\em Quantum}, 5:454, may 2021.

\bibitem{fuller2021approximate}
Bryce Fuller, Charles Hadfield, Jennifer~R. Glick, Takashi Imamichi, Toshinari
  Itoko, Richard~J. Thompson, Yang Jiao, Marna~M. Kagele, Adriana~W.
  Blom-Schieber, Rudy Raymond, and Antonio Mezzacapo.
\newblock Approximate solutions of combinatorial problems via quantum
  relaxations, 2021.

\bibitem{teramoto2023quantumrelaxation}
Kosei Teramoto, Rudy Raymond, Eyuri Wakakuwa, and Hiroshi Imai.
\newblock Quantum-relaxation based optimization algorithms: Theoretical
  extensions, 2023.

\bibitem{rancic2023}
Marko~J. Ran\ifmmode \check{c}\else \v{c}\fi{}i\ifmmode~\acute{c}\else
  \'{c}\fi{}.
\newblock Noisy intermediate-scale quantum computing algorithm for solving an
  $n$-vertex {M}ax{c}ut problem with log($n$) qubits.
\newblock {\em Phys. Rev. Res.}, 5:L012021, February 2023.

\bibitem{winderl2022comparative}
David Winderl, Nicola Franco, and Jeanette~Miriam Lorenz.
\newblock A comparative study on solving optimization problems with
  exponentially fewer qubits, 2022.

\bibitem{Liu_2019}
Jin-Guo Liu, Yi-Hong Zhang, Yuan Wan, and Lei Wang.
\newblock Variational quantum eigensolver with fewer qubits.
\newblock {\em Physical Review Research}, 1(2), September 2019.

\bibitem{decross2022qubitreuse}
Matthew DeCross, Eli Chertkov, Megan Kohagen, and Michael Foss-Feig.
\newblock Qubit-reuse compilation with mid-circuit measurement and reset, 2022.

\bibitem{hua2023exploiting}
Fei Hua, Yuwei Jin, Yanhao Chen, Suhas Vittal, Kevin Krsulich, Lev~S. Bishop,
  John Lapeyre, Ali Javadi-Abhari, and Eddy~Z. Zhang.
\newblock Exploiting qubit reuse through mid-circuit measurement and reset,
  2023.

\bibitem{lenstra1975some}
Jan~Karel Lenstra and AHG~Rinnooy Kan.
\newblock Some simple applications of the travelling salesman problem.
\newblock {\em Journal of the Operational Research Society}, 26(4):717--733,
  1975.

\bibitem{gavish1978travelling}
Bezalel Gavish and Stephen~C Graves.
\newblock The travelling salesman problem and related problems, 1978.

\bibitem{solomon1988survey}
Marius~M Solomon and Jacques Desrosiers.
\newblock Survey paper—time window constrained routing and scheduling
  problems.
\newblock {\em Transportation science}, 22(1):1--13, 1988.

\bibitem{solomon1987algorithms}
Marius~M Solomon.
\newblock Algorithms for the vehicle routing and scheduling problems with time
  window constraints.
\newblock {\em Operations research}, 35(2):254--265, 1987.

\bibitem{braysy2005vehicle}
Olli Br{\"a}ysy and Michel Gendreau.
\newblock Vehicle routing problem with time windows, part i: Route construction
  and local search algorithms.
\newblock {\em Transportation science}, 39(1):104--118, 2005.

\bibitem{kirkpatrick1983optimization}
Scott Kirkpatrick, C~Daniel Gelatt~Jr, and Mario~P Vecchi.
\newblock Optimization by simulated annealing.
\newblock {\em science}, 220(4598):671--680, 1983.

\bibitem{Rutenbar17235}
R.A. Rutenbar.
\newblock Simulated annealing algorithms: an overview.
\newblock {\em IEEE Circuits and Devices Magazine}, 5(1):19--26, 1989.

\bibitem{Bertsimas1993}
Dimitris Bertsimas and John Tsitsiklis.
\newblock {Simulated Annealing}.
\newblock {\em Statistical Science}, 8(1):10 -- 15, 1993.

\bibitem{Guilmeau9513782}
Thomas Guilmeau, Emilie Chouzenoux, and Víctor Elvira.
\newblock Simulated annealing: a review and a new scheme.
\newblock In {\em 2021 IEEE Statistical Signal Processing Workshop (SSP)},
  pages 101--105, 2021.

\bibitem{Romeo1991}
Fabio Romeo and Alberto Sangiovanni-Vincentelli.
\newblock A theoretical framework for simulated annealing.
\newblock {\em Algorithmica}, 6(1):302--345, Jun 1991.

\bibitem{EGLESE1990271}
R.W. Eglese.
\newblock Simulated annealing: A tool for operational research.
\newblock {\em European Journal of Operational Research}, 46(3):271--281, 1990.

\bibitem{glover1990tabu}
Fred Glover.
\newblock Tabu search: A tutorial.
\newblock {\em Interfaces}, 20(4):74--94, 1990.

\bibitem{Glover1998}
Fred Glover and Manuel Laguna.
\newblock Tabu search.
\newblock In Ding-Zhu Du and Panos~M. Pardalos, editors, {\em Handbook of
  Combinatorial Optimization: Volume1--3}, pages 2093--2229. Springer US,
  Boston, MA, 1998.

\bibitem{wang2012multilevel}
Yang Wang, Zhipeng L{\"u}, Fred Glover, and Jin-Kao Hao.
\newblock A multilevel algorithm for large unconstrained binary quadratic
  optimization.
\newblock In {\em Integration of AI and OR Techniques in Contraint Programming
  for Combinatorial Optimzation Problems: 9th International Conference, CPAIOR
  2012, Nantes, France, May 28--June1, 2012. Proceedings 9}, pages 395--408.
  Springer, 2012.

\bibitem{Fouskakis2002}
Dimitris Fouskakis and David Draper.
\newblock Stochastic optimization: a review.
\newblock {\em International Statistical Review}, 70(3):315--349, 2002.

\bibitem{goh2020hybrid}
Siong~Thye Goh, Sabrish Gopalakrishnan, Jianyuan Bo, and Hoong~Chuin Lau.
\newblock A hybrid framework using a qubo solver for permutation-based
  combinatorial optimization.
\newblock {\em arXiv preprint arXiv:2009.12767}, 2020.

\bibitem{math11010129}
Marcelo Becerra-Rozas, José Lemus-Romani, Felipe Cisternas-Caneo, Broderick
  Crawford, Ricardo Soto, Gino Astorga, Carlos Castro, and José García.
\newblock Continuous metaheuristics for binary optimization problems: An
  updated systematic literature review.
\newblock {\em Mathematics}, 11(1), 2023.

\bibitem{Doerr2021}
Benjamin Doerr and Frank Neumann.
\newblock A survey on recent progress in the theory of evolutionary algorithms
  for discrete optimization.
\newblock {\em ACM Trans. Evol. Learn. Optim.}, 1(4), oct 2021.

\bibitem{Dorigo6787854}
Marco Dorigo, Gianni~Di Caro, and Luca~M. Gambardella.
\newblock Ant algorithms for discrete optimization.
\newblock {\em Artificial Life}, 5(2):137--172, 1999.

\bibitem{Grama755612}
A.~Grama and V.~Kumar.
\newblock State of the art in parallel search techniques for discrete
  optimization problems.
\newblock {\em IEEE Transactions on Knowledge and Data Engineering},
  11(1):28--35, 1999.

\bibitem{Ricardo1996}
Ricardo Corr{\^e}a and Afonso Ferreira.
\newblock Parallel best-first branch- and-bound in discrete optimization: A
  framework.
\newblock In Afonso Ferreira and Panos Pardalos, editors, {\em Solving
  Combinatorial Optimization Problems in Parallel: Methods and Techniques},
  pages 171--200. Springer Berlin Heidelberg, Berlin, Heidelberg, 1996.

\bibitem{BLUM2005353}
Christian Blum.
\newblock Ant colony optimization: Introduction and recent trends.
\newblock {\em Physics of Life Reviews}, 2(4):353--373, 2005.

\bibitem{GHAREHCHOPOGH20191}
Farhad~Soleimanian Gharehchopogh and Hojjat Gholizadeh.
\newblock A comprehensive survey: Whale optimization algorithm and its
  applications.
\newblock {\em Swarm and Evolutionary Computation}, 48:1--24, 2019.

\bibitem{Wang2018}
Dongshu Wang, Dapei Tan, and Lei Liu.
\newblock Particle swarm optimization algorithm: an overview.
\newblock {\em Soft Computing}, 22(2):387--408, Jan 2018.

\bibitem{RizkAllah2023}
Rizk~M. Rizk-Allah and Aboul~Ella Hassanien.
\newblock A comprehensive survey on the sine--cosine optimization algorithm.
\newblock {\em Artificial Intelligence Review}, 56(6):4801--4858, Jun 2023.

\bibitem{KOULAMAS199441}
C~Koulamas, SR~Antony, and R~Jaen.
\newblock A survey of simulated annealing applications to operations research
  problems.
\newblock {\em Omega}, 22(1):41--56, 1994.

\bibitem{ELSHERBENY2010123}
Nasser~A. El-Sherbeny.
\newblock Vehicle routing with time windows: An overview of exact, heuristic
  and metaheuristic methods.
\newblock {\em Journal of King Saud University - Science}, 22(3):123--131,
  2010.

\bibitem{7047830}
Gitae Kim, Yew-Soon Ong, Chen~Kim Heng, Puay~Siew Tan, and Nengsheng~Allan
  Zhang.
\newblock City vehicle routing problem (city vrp): A review.
\newblock {\em IEEE Transactions on Intelligent Transportation Systems},
  16(4):1654--1666, 2015.

\bibitem{Doerner2010}
Karl~F. Doerner and Verena Schmid.
\newblock Survey: Matheuristics for rich vehicle routing problems.
\newblock In Mar{\'i}a~J. Blesa, Christian Blum, G{\"u}nther Raidl, Andrea
  Roli, and Michael Sampels, editors, {\em Hybrid Metaheuristics}, pages
  206--221, Berlin, Heidelberg, 2010. Springer Berlin Heidelberg.

\bibitem{MONTOYATORRES2015115}
Jairo~R. Montoya-Torres, Julián {López Franco}, Santiago {Nieto Isaza},
  Heriberto {Felizzola Jiménez}, and Nilson Herazo-Padilla.
\newblock A literature review on the vehicle routing problem with multiple
  depots.
\newblock {\em Computers \& Industrial Engineering}, 79:115--129, 2015.

\bibitem{Konstantakopoulos2022}
Grigorios~D. Konstantakopoulos, Sotiris~P. Gayialis, and Evripidis~P.
  Kechagias.
\newblock Vehicle routing problem and related algorithms for logistics
  distribution: a literature review and classification.
\newblock {\em Operational Research}, 22(3):2033--2062, Jul 2022.

\bibitem{Abualigah2020}
Laith Abualigah and Ali Diabat.
\newblock A comprehensive survey of the grasshopper optimization algorithm:
  results, variants, and applications.
\newblock {\em Neural Computing and Applications}, 32(19):15533--15556, Oct
  2020.

\bibitem{Cordeau2005}
Jean-Francois Cordeau and Gilbert Laporte.
\newblock Tabu search heuristics for the vehicle routing problem.
\newblock In Ramesh Sharda, Stefan Vo{\ss}, C{\'e}sar Rego, and Bahram Alidaee,
  editors, {\em Metaheuristic Optimization via Memory and Evolution: Tabu
  Search and Scatter Search}, pages 145--163. Springer US, Boston, MA, 2005.

\bibitem{basu2012tabu}
Sumanta Basu.
\newblock Tabu search implementation on traveling salesman problem and its
  variations: a literature survey.
\newblock {\em American Journal of Operations Research}, 2012.

\bibitem{Osaba9781399}
Eneko Osaba, Esther Villar-Rodriguez, and Izaskun Oregi.
\newblock A systematic literature review of quantum computing for routing
  problems.
\newblock {\em IEEE Access}, 10:55805--55817, 2022.

\bibitem{AQCReview2018}
Tameem Albash and Daniel~A. Lidar.
\newblock Adiabatic quantum computation.
\newblock {\em Rev. Mod. Phys.}, 90:015002, Jan 2018.

\bibitem{Yarkoni_2022}
Sheir Yarkoni, Elena Raponi, Thomas Bäck, and Sebastian Schmitt.
\newblock Quantum annealing for industry applications: introduction and review.
\newblock {\em Reports on Progress in Physics}, 85(10):104001, sep 2022.

\bibitem{Irie2019}
Hirotaka Irie, Goragot Wongpaisarnsin, Masayoshi Terabe, Akira Miki, and
  Shinichirou Taguchi.
\newblock Quantum annealing of vehicle routing problem with time, state and
  capacity.
\newblock In Sebastian Feld and Claudia Linnhoff-Popien, editors, {\em Quantum
  Technology and Optimization Problems}, pages 145--156, Cham, 2019. Springer
  International Publishing.

\bibitem{harikrishnakumar2020quantum}
Ramkumar Harikrishnakumar, Saideep Nannapaneni, Nam~H. Nguyen, James~E. Steck,
  and Elizabeth~C. Behrman.
\newblock A quantum annealing approach for dynamic multi-depot capacitated
  vehicle routing problem, 2020.

\bibitem{Borowski2020}
Micha{\l} Borowski, Pawe{\l} Gora, Katarzyna Karnas, Mateusz B{\l}ajda,
  Krystian Kr{\'o}l, Artur Matyjasek, Damian Burczyk, Miron Szewczyk, and
  Micha{\l} Kutwin.
\newblock New hybrid quantum annealing algorithms for solving vehicle routing
  problem.
\newblock In Valeria~V. Krzhizhanovskaya, G{\'a}bor Z{\'a}vodszky, Michael~H.
  Lees, Jack~J. Dongarra, Peter M.~A. Sloot, S{\'e}rgio Brissos, and Jo{\~a}o
  Teixeira, editors, {\em Computational Science -- ICCS 2020}, pages 546--561,
  Cham, 2020. Springer International Publishing.

\bibitem{Yarkoni2021}
Sheir Yarkoni, Andreas Huck, Hanno Sch{\"u}lldorf, Benjamin Speitkamp,
  Marc~Shakory Tabrizi, Martin Leib, Thomas B{\"a}ck, and Florian Neukart.
\newblock Solving the shipment rerouting problem with quantum optimization
  techniques.
\newblock In Martijn Mes, Eduardo Lalla-Ruiz, and Stefan Vo{\ss}, editors, {\em
  Computational Logistics}, pages 502--517, Cham, 2021. Springer International
  Publishing.

\bibitem{Azad9774961}
Utkarsh Azad, Bikash~K. Behera, Emad~A. Ahmed, Prasanta~K. Panigrahi, and Ahmed
  Farouk.
\newblock Solving vehicle routing problem using quantum approximate
  optimization algorithm.
\newblock {\em IEEE Transactions on Intelligent Transportation Systems},
  24(7):7564--7573, 2023.

\bibitem{fitzek2021applying}
David Fitzek, Toheed Ghandriz, Leo Laine, Mats Granath, and Anton~Frisk Kockum.
\newblock Applying quantum approximate optimization to the heterogeneous
  vehicle routing problem, 2021.

\bibitem{xie2023feasibilitypreserved}
Ningyi Xie, Xinwei Lee, Dongsheng Cai, Yoshiyuki Saito, Nobuyoshi Asai, and
  Hoong~Chuin Lau.
\newblock A feasibility-preserved quantum approximate solver for the
  capacitated vehicle routing problem, 2023.

\bibitem{Mohanty10214310}
Nishikanta Mohanty, Bikash~K. Behera, and Christopher Ferrie.
\newblock Analysis of the vehicle routing problem solved via hybrid quantum
  algorithms in the presence of noisy channels.
\newblock {\em IEEE Transactions on Quantum Engineering}, pages 1--16, 2023.

\bibitem{Alsaiyari10087522}
Muhammad Alsaiyari and Muhamad Felemban.
\newblock Variational quantum algorithms for solving vehicle routing problem.
\newblock In {\em 2023 International Conference on Smart Computing and
  Application (ICSCA)}, pages 1--4, 2023.

\bibitem{kossmann2022deep}
Gereon Ko{\ss}mann, Lennart Binkowski, Lauritz van Luijk, Timo Ziegler, and
  Ren{\'e} Schwonnek.
\newblock Deep-circuit qaoa.
\newblock {\em arXiv preprint arXiv:2210.12406}, 2022.

\bibitem{McClean2018}
Jarrod~R. McClean, Sergio Boixo, Vadim~N. Smelyanskiy, Ryan Babbush, and
  Hartmut Neven.
\newblock Barren plateaus in quantum neural network training landscapes.
\newblock {\em Nature Communications}, 9(1):4812, 2018.

\bibitem{cerezo2021cost}
Marco Cerezo, Akira Sone, Tyler Volkoff, Lukasz Cincio, and Patrick~J Coles.
\newblock Cost function dependent barren plateaus in shallow parametrized
  quantum circuits.
\newblock {\em Nature communications}, 12(1):1--12, 2021.

\bibitem{wang2021noise}
Samson Wang, Enrico Fontana, Marco Cerezo, Kunal Sharma, Akira Sone, Lukasz
  Cincio, and Patrick~J Coles.
\newblock Noise-induced barren plateaus in variational quantum algorithms.
\newblock {\em Nature communications}, 12(1):6961, 2021.

\bibitem{bharti2020quantum}
Kishor Bharti.
\newblock Quantum assisted eigensolver.
\newblock {\em arXiv preprint arXiv:2009.11001}, 2020.

\bibitem{bharti2021iterative}
Kishor Bharti and Tobias Haug.
\newblock Iterative quantum-assisted eigensolver.
\newblock {\em Physical Review A}, 104(5):L050401, 2021.

\bibitem{zhang2018improved}
Zijun Zhang.
\newblock Improved adam optimizer for deep neural networks.
\newblock In {\em 2018 IEEE/ACM 26th International Symposium on Quality of
  Service (IWQoS)}, pages 1--2. Ieee, 2018.

\bibitem{crooks2019gradients}
Gavin~E Crooks.
\newblock Gradients of parameterized quantum gates using the parameter-shift
  rule and gate decomposition.
\newblock {\em arXiv preprint arXiv:1905.13311}, 2019.

\bibitem{bergholm2022pennylane}
Ville Bergholm, Josh Izaac, Maria Schuld, Christian Gogolin, Carsten Blank,
  Keri McKiernan, and Nathan Killoran.
\newblock Pennylane: Automatic differentiation of hybrid quantum-classical
  computations, 2022.

\bibitem{ibmq}
IBM~Quantum team. Retrieved from~https://quantum computing.ibm.com.
\newblock Ibmq, 2020.

\bibitem{braket}
Amazon~Braket. https://aws.amazon.com/braket/.

\bibitem{ionq}
IonQ Trapped Ion Quantum~Computing https://ionq.com/.

\bibitem{Gurobi}
LLC Gurobi~Optimization.
\newblock Gurobi optimizer reference manual, 2020.

\end{thebibliography}

\end{document}